\documentclass[
  aps, 
 superscriptaddress,
 pre,
 twocolumn,
 showpacs,
 amsmath,amssymb,
]{revtex4}

 \usepackage{xcolor}
 \usepackage{color}
\usepackage{hyperref}
\hypersetup{
    urlbordercolor=black, 
  colorlinks   = true, 
  urlcolor     = black, 
  linkcolor    = black, 
  citecolor   = black 
}

\usepackage{graphicx}
\usepackage{dcolumn}

\usepackage{bm}
\usepackage{color}
\usepackage{dcolumn}
\newcolumntype{M}[1]{>{\centering\arraybackslash}m{#1}}

\newcommand{\mc}[1]{\multicolumn{1}{c}{#1}}

\begin{document}

\preprint{APS/123-QED}

\title{Anisotropic confinement effects in a two-dimensional plasma crystal}

\author{I.~Laut}
\email{Ingo.Laut@dlr.de}
\affiliation{Deutsches Zentrum f\"ur Luft- und Raumfahrt, Forschungsgruppe Komplexe Plasmen, 82234 We{\ss}ling, Germany}

\author{S. K.~Zhdanov}
\affiliation{Max Planck Institute for extraterrestrial Physics, 85741 Garching, Germany}

\author{C.~R\"{a}th}
\affiliation{Deutsches Zentrum f\"ur Luft- und Raumfahrt, Forschungsgruppe Komplexe Plasmen, 82234 We{\ss}ling, Germany}

\author{H. M.~Thomas}
\affiliation{Deutsches Zentrum f\"ur Luft- und Raumfahrt, Forschungsgruppe Komplexe Plasmen, 82234 We{\ss}ling, Germany}

\author{G. E.~Morfill}
\affiliation{Max Planck Institute for extraterrestrial Physics, 85741 Garching, Germany}
\affiliation{BMSTU Centre for Plasma Science and Technology, Moscow, Russia}

\date{\today}

\begin{abstract}
{
The spectral asymmetry of the wave energy distribution of dust particles during mode-coupling induced melting, observed for the first time in plasma crystals by Cou\"edel \emph{et al.} [Phys.\ Rev.\ E {\bf89}, 053108 (2014)], is studied theoretically and by molecular-dynamics simulations. It is shown that an anisotropy of the well confining the microparticles selects the directions of preferred particle motion. The observed differences in intensity of waves of opposed directions is explained by a nonvanishing phonon flux. Anisotropic phonon scattering by defects and Umklapp scattering are proposed as possible reasons for the mean phonon flux. 
}
\end{abstract}

\maketitle

\section{Introduction}
\label{sec_introduction}

Complex or dusty plasmas are weakly ionized gases containing micron-size particles. In a laboratory radio-frequency (rf) plasma, these particles are negatively charged and thus repel each other. In rf discharge complex plasmas, the particles are self-trapped in the plasma \cite{chu1994, thomas1994, hayashi1994}. 
Due to their strong interactions with the plasma and with each other, they can form strongly coupled crystals \cite{fortov2005, morfill2009}, called \emph{plasma crystals}. Complex plasmas are ideal model systems for phase transitions \cite{thomas1996, schweigert1998}, transport processes \cite{nunomura2000, misawa2001, nunomura2005heat, nosenko2008heat} and self-organization \cite{menzel2010, williams2014}. In ground-based experiments, the particles levitate in the plasma sheath region above the lower electrode where they can form a horizontal two-dimensional (2D) monolayer under adequate experimental conditions \cite{chu1994, thomas1994, hayashi1994}. Due to the finite vertical confinement of the crystal, the monolayer is not completely flat, allowing an out-of-plane wave mode which has an optical dispersion relation in addition to the two in-plane modes with acoustic dispersion. 

The surrounding plasma strongly influences the particle-particle interaction, making it anisotropic. While the mutual repulsion of equally-charged particles is ascribed to a Yukawa potential \cite{ikezi1986}, an attractive component stems from the plasma wake \cite{melzer1999} which is formed beneath every particle downstream of the ion flow. In theory and simulations, the plasma wake is often modeled as a pointlike effective charge below each particle \cite{ivlev2000}. If the vertical particle confinement is small enough, a \emph{mode-coupling instability} (MCI) can occur in such a model, coupling the out-of-plane mode to the longitudinal mode \cite{ivlev2000, zhdanov2009}. Near the intersection of the modes, the unstable hybrid mode grows until the crystalline order breaks. The experimental observations are in very good agreement with the predictions of the model \cite{couedel2010, couedel2011}. 

In an ideal hexagonal lattice, the MCI is equally strong in all three main directions of the crystal, reflecting its sixfold symmetry \cite{couedel2011}. In Ref.~\cite{couedel2014}, however, the instability was well pronounced dominantly in only one direction. A synchronization pattern of alternating in-phase and anti-phase oscillations accompanied the asymmetric triggering of MCI. Similar symmetry-breaking patterns were observed in colloids on global \cite{reichhardt2004} and intermediate \cite{bohlein2012} scales. A lattice deformation was suggested in Ref.~\cite{couedel2014} as a possible explanation for the symmetry breaking, though it was experimentally difficult to study. It was shown in simulations that the asymmetry of MCI can be caused by an anisotropy of the horizontal confinement \cite{laut2015}. Under adequate conditions, the instability can be active only in the direction of the compression of the crystal. The main conclusion was that for an appropriate orientation of the anisotropy, MCI could be triggered in two directions which leads to competing synchronization patterns. If MCI is triggered in one direction, a single pattern dominates. It was not possible, however, to explain a left-right asymmetry of opposed directions which was also present in the spectra \cite{laut2015}. 

In theoretical treatments the presence of the \emph{finite} horizontal confinement of the crystal is often ignored. For doing so there are certain arguments in addition to facilitating the theoretical description: (i) The horizontal confinement is known to be 100--200 times weaker than the vertical confinement \cite{samsonov2005}, allowing systems that are very extended in the horizontal direction. (ii) 'Confinement-free' systems (so called Yukawa systems) of mutually repelling particles are an excellent substitute to explain many, sometimes very delicate effects observed in experiments. (iii) The results obtained seem to be universal and important for many applications.

Still, the simplification of an infinite plasma crystal is not always justified. In the problem considered here the finite confinement is explicitly taken into account. The competition of the sixfold symmetry of the crystal lattice and the radial symmetry of the horizontal confinement leads to defects and inhomogeneities in the crystal. The actual configuration of the confining fields affects the structure of the microparticle cloud. 

In this paper we would like to highlight and report on the physics of \emph{spontaneous breaking of spectral symmetry} of the wave energy distribution of an anisotropically confined plasma crystal during the early stage of MCI. This asymmetry plays an eminent role in the understanding of the synchronization processes observed in experiments \cite{couedel2014} and simulations \cite{laut2015}, and may give hints to the connection to the recently discovered \emph{chimera states} that further fueled the interest in oscillator networks with controllable eigenfrequencies, coupling and topology \cite{kuramoto2002, abrams2004, motter2010, hagerstrom2012}.

The paper is organized as follows. In Sec.~\ref{sec_simulation_particlulars}, the numerical algorithm and the simulation procedure are described. In Sec.~\ref{sec_results}, the spectral asymmetry of a simulated complex plasma crystal is analyzed and compared to a theoretical model. In Sec.~\ref{sec_possible_origin_of_symmetry_braking}, the origin of this symmetry breaking is investigated in detail. The anisotropic phonon scattering by defects and the anisotropic Umklapp scattering are identified as two possible mechanisms introducing the asymmetry. Finally, in Sec.~\ref{sec_conclusion}, we conclude with a summary and discussion of our results.

\section{Simulation particulars}
\label{sec_simulation_particlulars}

\subsection{Governing equations} 
Molecular-dynamics simulations have proven to be an adequate tool to study and compare a wide range of experimental conditions. In the simulations, the potential well that confines the particles is treated as a tunable parameter, allowing to control the lattice configuration \cite{totsuji2001, zhdanov2011spontaneous}, crystal stability \cite{ivlev2003, rocker2014effect}, and anisotropy effects \cite{laut2015}. A parabolic confinement well is often used to simulate a monolayer suspension \cite{totsuji2001, zhdanov2011spontaneous, zhdanov2003large, sheridan2008, sheridan2009}. 
To model a monolayer extended in the $xy$ plane, a highly anisotropic three-dimensional confinement well, about 100 times stronger vertically than horizontally, is used  \cite{ivlev2003, rocker2014effect, laut2015}. In addition, the horizontal confinement can easily be made anisotropic as is explained below. 

The equations of motion employed in simulations read \cite{couedel2011, laut2015}:
\begin{equation}\label{eq_EqMotion}
M\ddot{\mathbf{r}}_i + M \nu \dot{\mathbf{r}}_i = \sum_{j \neq i} \mathbf{F}_{ji} + \mathbf{C}_i + \mathbf{L}_i,
\end{equation}
where $\mathbf{r}_i$ is the position of the $i$th particle
($i=1\ldots N$, $N$ the total number of particles),
$M $ the particle mass and $\nu $ the damping rate. The particle dynamics are governed by the mutual particle-particle interactions ($\mathbf{F}_{ji}$), the external interactions which are enabling confinement of the particle cloud ($\mathbf{C}_i$), and a heat bath ($\mathbf{L}_i$).

\begin{figure}
\includegraphics[width=\columnwidth]{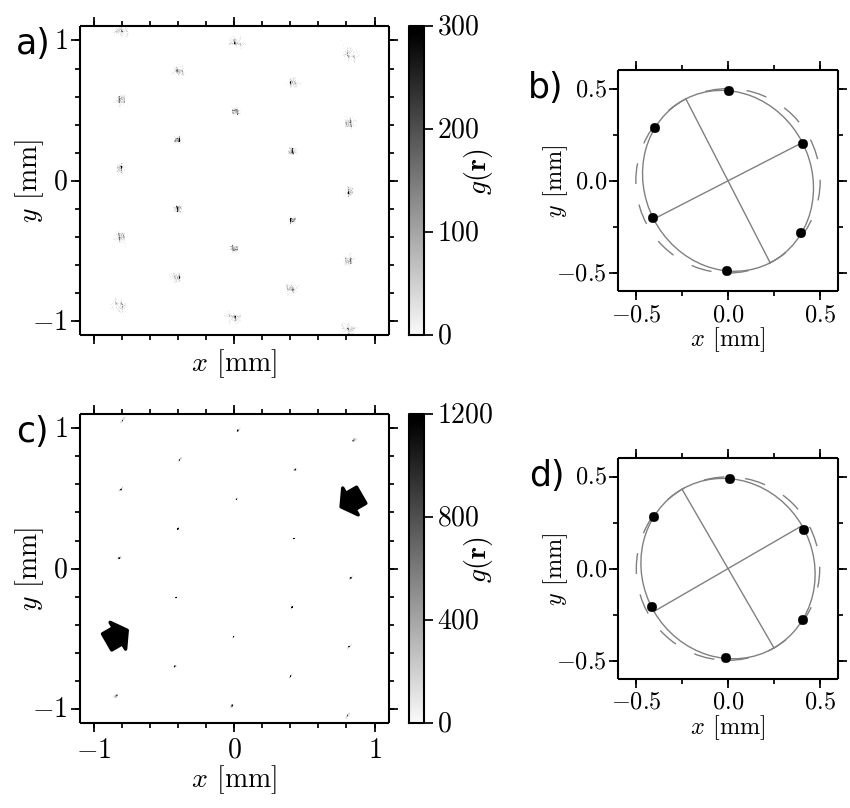}
\caption{
Pair correlation $g(\mathbf{r})$ for horizontally compressed crystals.
(a)~$g(\mathbf{r})$ in the horizontal plane of the experimental data of Ref.~\cite{couedel2014}. A central square of side length $13$\,mm was considered at time $t=0$.
(b)~The first peaks of $g(\mathbf{r})$ shown as solid circles. An ellipse (solid line) is fitted to the positions of the peaks; its deviation from a circle (dashed line) can be clearly seen.
(c),~(d): The same for the simulated data of Run~I (see Table~\ref{tab_1_params}). Here, $t=0$ corresponds to the starting point of the Dynamical phase. In panel (c), the direction of the angle of compression $\alpha$ is indicated by black arrows.
} \label{fig_1_pc}
\end{figure}

To characterize the particle confinement, it is instructive to introduce the (isotropic) horizontal confinement parameter $\Omega_c$ as well as the strength~$p$ and the direction~$\alpha$ of the loading asymmetry. The strength of the vertical confinement is characterized by $\Omega_z$. The external confinement $\mathbf{C}_i=\{C_{i,x},C_{i,y},C_{i,z}\}$ then reads (cf.~\cite{laut2015})
\begin{equation}\label{eq_ConForce}
 \begin{gathered}
  C_{i, x} = - M \Omega_c^2 X_i,\, C_{i, y} = - M \Omega_c^2 Y_i,\, C_{i, z}= - M \Omega_z^2 z_i, \\  
  X_i = x_i + p \left( x_i \cos 2\alpha + y_i \sin 2\alpha \right), \\
  Y_i = x_i + p \left( x_i \sin 2\alpha - y_i \cos 2\alpha \right).
 \end{gathered}
\end{equation}
Horizontally, the simulated crystal can thus be dominantly compressed under any angle $\alpha$ measured from the $x$ axis (see Fig.~\ref{fig_1_pc}). For instance, at $\alpha=0$ the horizontal confinement forces are distributed so that $C_{i ,x}= - m \Omega_c^2 x_i (1+p),\,C_{i ,y}= - m \Omega_c^2 y_i (1-p)$, and therefore the confinement is $(1+p)/(1-p)$ times 'stronger' in $x$ direction than in $y$ direction. It is also useful to define the confinement frequencies parallel $2 \pi f_\| = \Omega_c \sqrt{1 + p}$ and perpendicular $2 \pi f_\perp = \Omega_c \sqrt{1 - p}$ to the direction of the compression denoted by angle $\alpha$.  

Following Refs.~\cite{couedel2014, laut2015}, the force exerted by particle $j$ (and its wake) on particle $i$ is introduced as
\begin{equation}
\begin{split}
\mathbf{F}_{ji} =
 & \frac{Q^2}{r^2_{ji}} \exp \left( -\frac{r_{ji}}{\lambda} \right) \left( 1 + \frac{r_{ji}}{\lambda} \right) \frac{\mathbf{r}_{ji}}{r_{ji}} \\
 & -    \frac{q|Q|}{r^2_{w_{ji}}} \exp \left( -\frac{r_{w_{ji}}}{\lambda} \right) \left( 1 + \frac{r_{w_{ji}}}{\lambda} \right) \frac{\mathbf{r}_{w_{ji}}}{r_{w_{ji}}},
\end{split}
\end{equation}
where $Q<0$ is the particle charge, $\lambda$ is the screening length, 
$\mathbf{r}_{ji} = \mathbf{r}_{i} - \mathbf{r}_{j}$ and 
$\mathbf{r}_{w_{ji}} = \mathbf{r}_{i} - (\mathbf{r}_{j} - \delta \mathbf{e}_z)$, where $\mathbf{e}_z$ is the ('vertical') unit vector perpendicular to the monolayer plane. The pointlike wake charge $q$ ($0<q < |Q|$) is located at a distance $\delta$ ($\delta < \lambda$) below each particle. The particle charges, the screening length and the wake parameters are considered as fixed in every simulation run (see Table~\ref{tab_1_params}).

\begin{table} 
 \caption{Parameters of the simulation runs. }
 \label{tab_1_params}
 \begin{ruledtabular} 
  \begin{tabular}{lcc}
 Parameter  & \text{Run I}    & \text{Run II} \\
\colrule\noalign{\smallskip}
 $N$                 & 16384    & 10000         \\
 $M$ (pg)            & 610      & 610           \\
 $Q$ ($e$)           & $-19000$ & $-19000$    \\
 $\nu$ (s$^{-1}$)    & 1.26     & 1.26          \\
 $\lambda$ ($\mu$m)  & 380      & 380           \\
 $q/|Q|$             & 0.2      & 0.2           \\
 $\delta/\lambda$    & 0.3      & 0.3           \\
 $\alpha$ ($^\circ$) & 30       & 0             \\
\colrule\noalign{\smallskip}
                                                 \multicolumn{3}{c}{Equilibration phase} \\
\colrule\noalign{\smallskip}
 $f_z$ (Hz)          & 23.0     & 22.0 \\
 $f_{\|}$ (Hz)       & 0.145    & 0.19 \\
 $f_{\bot}$ (Hz)     & 0.145    & 0.19 \\
\colrule\noalign{\smallskip}
                                                 \multicolumn{3}{c}{Deformation phase} \\
\colrule\noalign{\smallskip}
 $f_z$ (Hz)          & 23.0     & 22.0 \\
 $f_{\|}$ (Hz)       & 0.156    & 0.20 \\
 $f_{\bot}$ (Hz)     & 0.137    & 0.18 \\
\colrule\noalign{\smallskip}
                                                 \multicolumn{3}{c}{Dynamical phase} \\
\colrule\noalign{\smallskip}
 $f_z$ (Hz)          & 20.0     & 19.5 \\
 $f_{\|}$ (Hz)       & 0.156    & 0.20 \\
 $f_{\bot}$ (Hz)     & 0.137    & 0.18 \\
\end{tabular}
\end{ruledtabular}
\end{table} 

The particles are also coupled to a Langevin heat bath of temperature $T = 300$\,K,
\begin{equation}
\langle \mathbf{L}_i(t) \rangle = 0, \hspace{0.3cm} \langle \mathbf{L}_i(t + \tau) \mathbf{L}_j(t) \rangle = 2 \nu m T \delta_{ij} \delta(\tau).
\end{equation}
$\delta_{ij}$ is the Kronecker delta and $\delta(\tau)$ is the delta function. It is a commonly used approximation that allows one to simulate the random excitations stemming from the gas surrounding the particles \cite{sheridan2008, sheridan2009} (or plasma, as necessary \cite{zhdanov2011spontaneous}).

\subsection{Simulation procedure} 

The equations of motion (\ref{eq_EqMotion}) were integrated using the Beeman algorithm with predictor-corrector modifications \cite{schofield1973, beeman1976}. The code is parallelized using OpenMP. The vertical confinement frequency was about two orders of magnitude larger than the horizontal confinement frequencies, leading to the formation of quasi-2D monolayers.

Every simulation run was divided into three main phases characterized by three confinement frequencies each, see Table~\ref{tab_1_params}. The particles were initially positioned on a hexagonal grid. During the Equilibration phase, which is characterized by an anisotropic horizontal confinement and a large vertical confinement that prevents the onset of MCI, the crystal was allowed to relax. The competition of the hexagonal symmetry and the radial confinement lead to the melting of the outer region of the crystal, which then recrystallized to different domains divided by strings of defects. The central region kept the crystal structure. After equilibration, during the Deformation phase, the horizontal confinement well was modified to a desirable anisotropic configuration while the strong vertical confinement was kept untouched, and the particle cloud was allowed to relax further. Finally, after reaching the stable deformed configuration, the vertical confinement was reduced in the Dynamical phase to trigger the MCI. The temperature of the heat bath and all other parameters were fixed in the simulation runs.

\section{Results}
\label{sec_results}

\subsection{Pair correlations under loading asymmetry} 

In simulation Run~I (see Table~\ref{tab_1_params}), a monolayer of 16384 particles, each with a mass of $M = 6.1 \times 10^{-13}~\text{kg}$, was formed during the Equilibration phase at $f_\parallel = f_\perp = 0.145$\,Hz and $f_z = 23$\,Hz. The horizontal frequencies were then changed to $f_\parallel = 0.156$\,Hz parallel to direction $\alpha = 30^\circ$ and $f_\perp = 0.137$\,Hz perpendicular to it in order to introduce an anisotropy corresponding to the loading asymmetry of about $p=14\%$. Finally, in the Dynamical phase, the vertical confinement was reduced to $f_z = 20$\,Hz in order to start the instability, see Ref.~\cite{laut2015} for details.

The particle positions are analyzed in a window containing about 800 particles near the center of the crystal that showed synchronized motion. Both in experiments and in simulations the first peaks of the radial pair correlation function $g(r)$ are split in two compared to the expected peaks for an ideal hexagonal lattice \cite{couedel2014, laut2015}. Indeed, from the 2D pair correlation function $g(\mathbf{r})$ (see Fig.~\ref{fig_1_pc}) it can be seen that the distance to the nearest neighbors is about $7$\,\% smaller under an angle of $30^\circ$ than in the other two directions. The good agreement of experiment and simulation demonstrates that the asymmetry of $g(\mathbf{r})$ can be attributed to an anisotropic compression of the crystal in the horizontal plane.

\subsection{Asymmetric energy distribution} 

\begin{figure}
\centering
\includegraphics[width=\columnwidth]{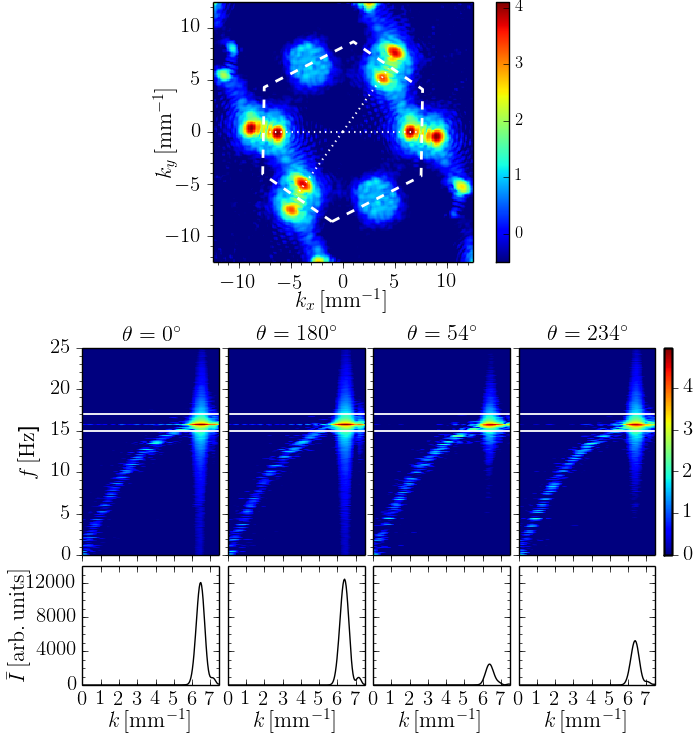}
\caption{
(Color online) Asymmetry of the MCI for compressed crystals.
Top panel: Intensity $\bar I_\mathbf{k}$ of the velocity fluctuation spectrum in the $k_x k_y$ plane for the simulated data of Run~I (see Table~\ref{tab_1_params}), averaged over the interval $15\,\mathrm{Hz} < f < 17\,\mathrm{Hz}$. The white dashed line indicates the border of the first Brillouin zone. The dotted lines with angles of $\theta = 0^\circ,\,54^\circ$ measured from the $x$~axis are shown to emphasize the symmetry of the hot-dot locations. 
Middle panels: Velocity fluctuation spectra as a function of $k$ and $f$ in the directions $\theta = 0^\circ,\,180^\circ,\,54^\circ,\,234^\circ$. 
Bottom panels: Intensity of the fluctuation spectra shown in the middle panels, averaged over the range $15\,\mathrm{Hz} < f < 17\,\mathrm{Hz}$ indicated by the horizontal lines. To compute the spectra, the first $25\,\mathrm{s}$ of the Dynamical phase were used in a central region of the crystal containing about 800 particles. Only longitudinal modes are shown. The colorbars are in a logarithmic scale with base $10$ in arbitrary units.  
} \label{fig_2_spectra_run1}
\end{figure}

The distribution of the fluctuation energy of the simulated crystal (as well as in experiments \cite{couedel2014}) is dominated by the hot dots (HDs), moreover it is highly asymmetric. To visualize the intensity of the particle current fluctuations $I_{\mathbf{k}, \omega}$, it is instructive to average over a frequency range around the hybrid frequency of the coupled longitudinal and transversal modes $f_\text{hyb} = (16 \pm 1)$\,Hz. The 2D map of the averaged intensity $\bar I_\mathbf{k}$ in the $k_xk_y$ plane is shown in Fig.~\ref{fig_2_spectra_run1}, top panel. As can be seen in this map, HDs are apparent in only two of the three main directions of the hexagonal lattice. The current fluctuation spectra $I_{\mathbf{k}, \omega}$ are calculated from the Fourier transform of the particle currents \cite{donko2008, couedel2014}. The border of the first Brillouin zone (fBz) is calculated from the static structure factor $S(\mathbf{k}) = N^{-1} \langle \sum_{l, m} e^{i \mathbf{k} \cdot (\mathbf{r}_l - \mathbf{r}_m)}\rangle$, where the sum runs over all pairs of particles, and the averaging is performed over time. The HDs appear as regions of high intensity inside the fBz. 

Note that the HDs at $\theta \simeq 0^\circ$ and $\theta \simeq 180^\circ$ are slightly brighter than the HDs at $\theta\simeq 54^\circ$ and $\theta \simeq 234^\circ$, see Fig.~\ref{fig_2_spectra_run1}, top panel. To further study the anisotropy of the fluctuation spectra, in the middle panels of Fig.~\ref{fig_2_spectra_run1} the intensity $I_{\mathbf{k}, \omega}$ is shown as a function of the modulus of the wave vector $k$ and of frequency $f$ in those directions. By averaging the one-dimensional spectra over the frequency range of interest (indicated as horizontal lines in the Figure) one can compare the intensities of the peaks, see Fig.~\ref{fig_2_spectra_run1}, bottom panels. It becomes apparent that the HD intensities in the direction of the $x$ axis ($\theta = 0^\circ,\,180^\circ$) are more than a factor of two stronger than the intensities of the HDs in the other direction ($\theta = 54^\circ,\, 234^\circ$).

Also note that while the HDs on the $x$ axis are nearly equally bright, the HDs at $\theta = 54^\circ$ and $\theta = 234^\circ$ are highly asymmetric, see Fig.~\ref{fig_2_spectra_run1}, bottom panels. Strictly speaking, the HD energy distribution is neither mirror nor rotationally symmetric, indicating strong symmetry breaking. All these results are in a very good qualitative agreement with experimental observations \cite{couedel2014, laut2015}. The character of asymmetry indicates the presence of the dominant phonon flux in the $\approx234^\circ$ direction; see section \ref{subsec_flux} below.

\subsection{Interaction range of the confined crystal} 

The anisotropy of the spectral intensity of the particle velocity fluctuations caused by the weakly angle-dependent loading indicates that the MCI is sensitive to a variation of the confinement strength \cite{laut2015}. The horizontal confinement of the crystal is often assumed to be insignificant in theoretical considerations, see, e.g., \cite{zhdanov2009, ivlev2015wave}. On the contrary, the finiteness and symmetry of the confinement have a great influence on the delicate symmetry breaking effects.

The cluster density and its spatial distribution varies with the strength of the horizontal confinement $\Omega_c$, making the particle cluster internally inhomogeneous. It is not difficult to examine the character of this deformation. The confinement technique implemented in the simulations, caging the particle cluster in a parabolic potential, is actually well known, as well as the scaling laws controlling the structure of such Yukawa-interacting particle clusters, see, e.g., \cite{durniak2011, durniak2013, totsuji2001} and the references therein. According to \cite{peeters1987, totsuji2001, zhdanov2011spontaneous} at fixed particle charge ($Q$), screening length ($\lambda$), and number of particles ($N$), the following approximate relationships hold
\begin{equation}\label{eq_SL}
\frac{c_l^2}{\Omega_c^2Ra}\propto\frac{a}{R}\propto\left(\frac{\kappa c_l}{c_q}\right)^2\simeq \text{const},
\end{equation}
where $c_l$ is the longitudinal sound speed, $R$ the cluster size, $a$ the crystal constant, $\kappa=a/\lambda$ the interaction range, $c_q=|Q|/\sqrt{M\lambda}$, and a 'const' to the right means a function that rather weakly depends on $\kappa$. To the same accuracy, from Eq.~(\ref{eq_SL}) it follows immediately that
\begin{equation}\label{eq_kappa}
\Omega_c\kappa^2\simeq \text{const},
\end{equation}
and the direct dependence of the cluster interaction range $\kappa$ on the confinement strength becomes apparent. The large scale density distribution is readily studied more rigorously, in analogy to \cite{totsuji2001, dubin1997}, by minimizing the cluster interaction energy; see Appendix~\ref{app1} for details.

\begin{figure}
\includegraphics[width=.4\linewidth, trim=0 1.75cm 0 0, clip]{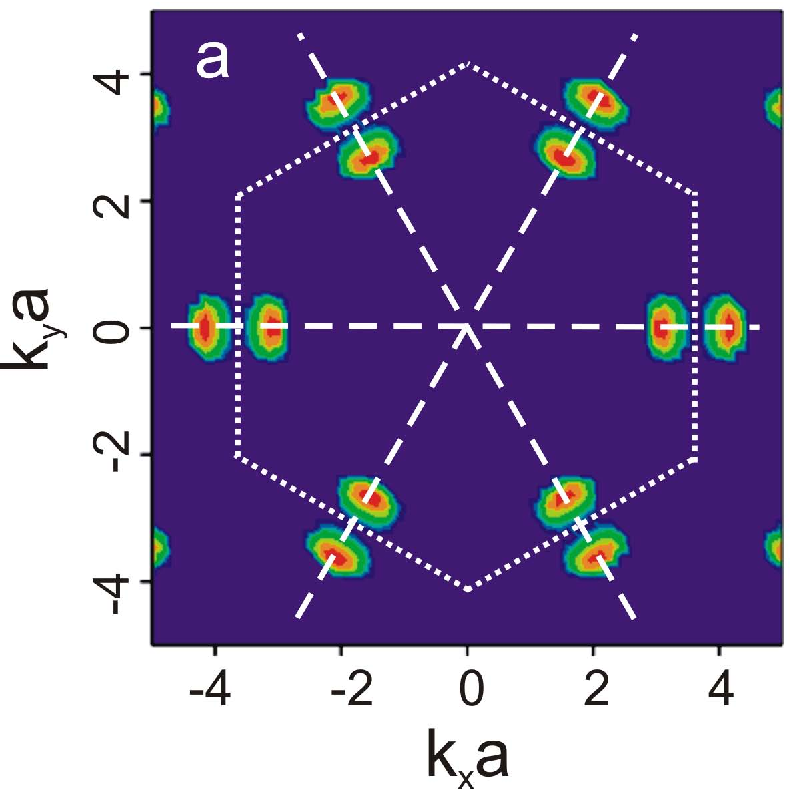}
\includegraphics[width=.65\linewidth, trim=0 1.75cm 0 0, clip]{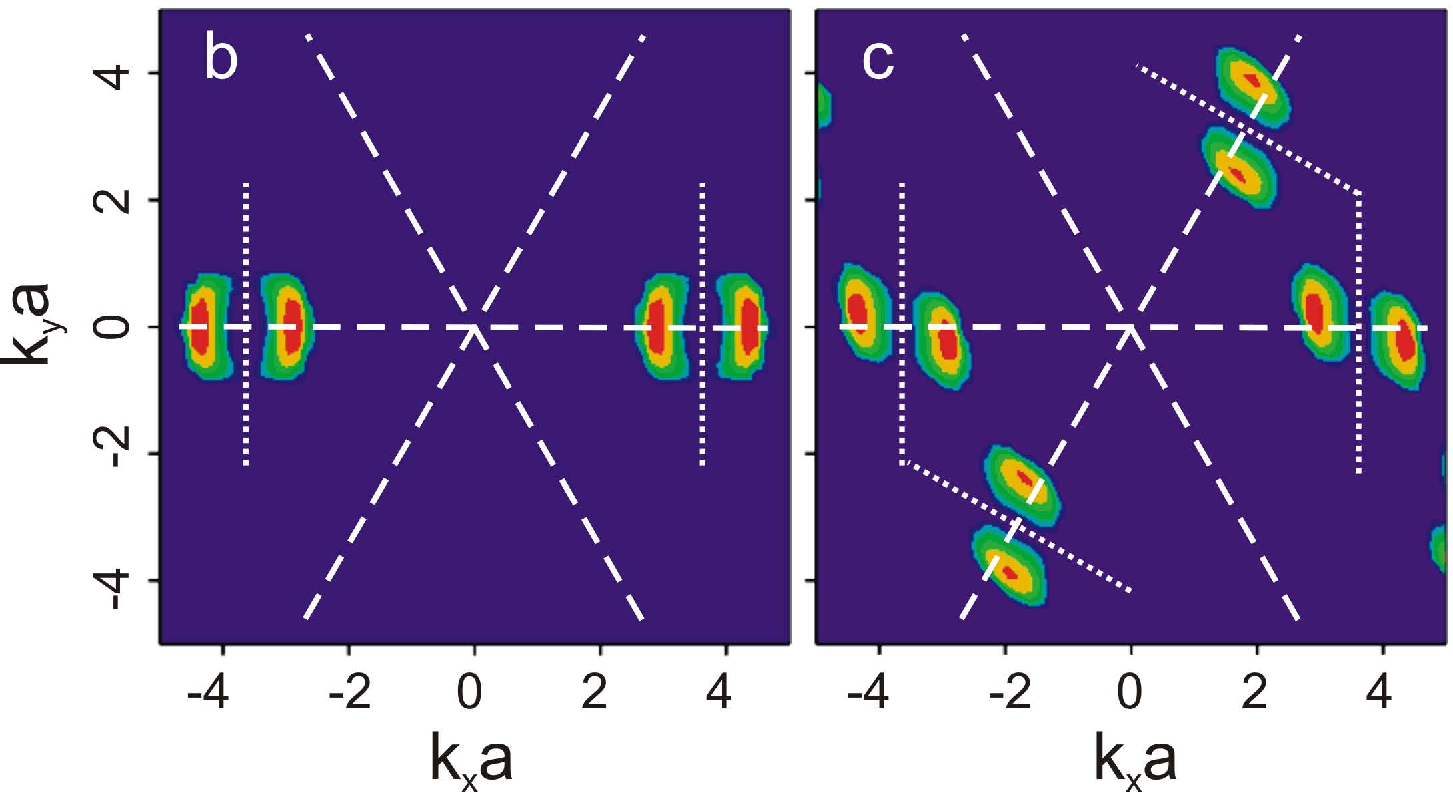}
\includegraphics[width=.65\linewidth]{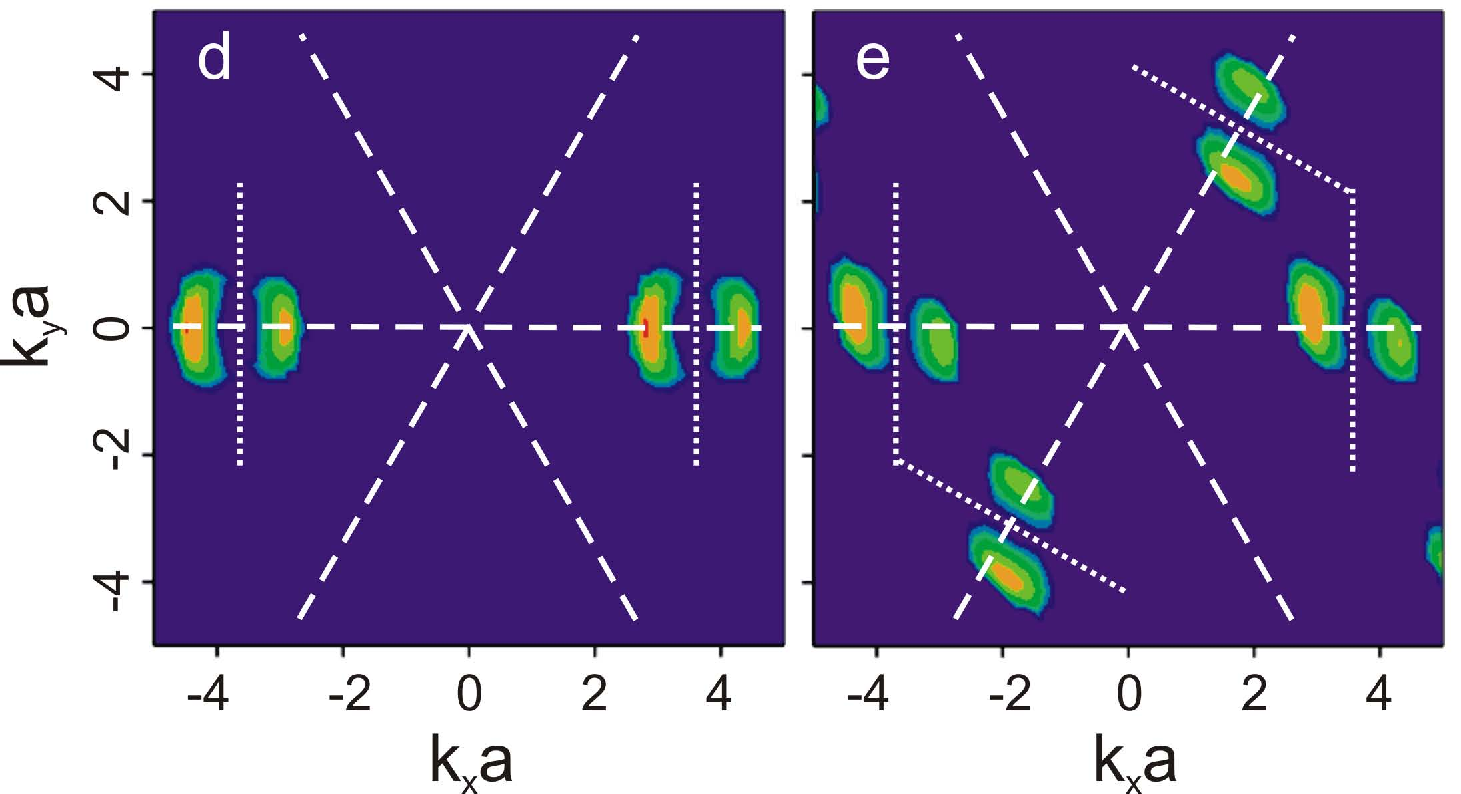}
\caption{
(Color online) Symmetry of the MCI increment in reduced $k_xk_y$ maps calculated for the theoretical model.
In (a), the increment is calculated for isotropic loading as in Ref.~\cite{zhdanov2009}. 
In (b) and (c), there is an anisotropic loading of strength $p=14$\% in the direction $\alpha=0^\circ$ and $\alpha=30^\circ$, respectively. 
Panels (d) and (e) show the same as (b) and (c), only with a nonzero phonon flux in $x$ direction. See Table~\ref{tab_2_theory_params} for the parameters of the calculation. The dashed and dotted lines indicate the main crystallographic directions and the first Brillouin zone boundary of the unperturbed lattice. 
} \label{fig_3_spectra_theory}
\end{figure}

\begin{table} 
 \caption {Asymmetry of the theoretical spectra. The lines correspond to the different panels in Fig.~\ref{fig_3_spectra_theory}. The parameters $\Omega_z/\Omega_Q$ and $\alpha$ were used to calculate the spectra. The wave group-to-phase velocity ratio at the hot dots $u/u_\text{HD}$, the growth rate ratio of the twin hot dots $\delta \gamma / \langle \gamma \rangle$ and the hot-spot positions $\langle k_\text{HD}a \rangle$ are calculated for the angles $\theta$ specified in the fourth row. The MCI increments are computed assuming $\delta/\lambda=0.33$, $q/|Q|=0.3$, $\kappa=1$. Designations: 
 $\Omega_Q=|Q|/\sqrt{M\lambda^3}$, 
 $\delta\gamma/\langle\gamma\rangle=2(\gamma_{\theta}-\gamma_{\theta+\pi})/(\gamma_{\theta}+\gamma_{\theta+\pi})$.
 } \label{tab_2_theory_params}
 \begin{ruledtabular}
  \begin{tabular}{c d c c c c c}
   Panel & \mc{$\Omega_z/\Omega_Q$} &$\alpha$    &$\theta$     & $u/u_\text{HD}$ & $\delta\gamma/\langle\gamma\rangle$ & $\langle k_\text{HD}\rangle a$ \\      
   \colrule\noalign{\smallskip}
   a     & 4.15                     & 0          & 0           & 0               & 0                                   & 3.06                         \\ 
   b     & 4.2                      & 0          & 0           & 0               & 0                                   & 2.86                         \\ 
   c     & 4.2                      & $\pi/6$    & 0           & 0               & 0                                   & 2.94                         \\ 
   d     & 4.2                      & 0          & 0           & 0.28            & 0.09                                & 2.87                         \\ 
   e     & 4.2                      & $\pi/6$    & 0           & 0.25            & 0.17                                & 2.94                         \\ 
   e     & 4.2                      & $\pi/6$    & $\pi/3$     & 0.32            & 0.09                                & 2.98                         \\ 
  \end{tabular}
 \end{ruledtabular}
\end{table}

Since the MCI threshold critically depends on the crystal interaction range $\kappa$ \cite{couedel2011}, relationship (\ref{eq_kappa}) makes the critical vertical confinement  $ \Omega_{z,\text{crit}}$ (below which the instability is triggered) directly dependent on the horizontal confinement strength. It has been shown in Ref.~\cite{laut2015} that
\begin{equation}\label{eq_FirstVariation}
\frac{\delta\Omega_{z,\text{crit}}}{\Omega_{z,\text{crit}}}\approx\frac{\delta\Omega_c}{\Omega_c}.
\end{equation}

Under the anisotropic loading, the horizontal confinement strength  as well as the crystal interaction range are angle-dependent (elliptic-shaped, see Appendix~\ref{app2}), and, as a consequence, the MCI ignition becomes anisotropic. Given the angular dependence of the crystal structure is rather weak (see Fig.~\ref{fig_1_pc}), the spectral anisotropy of the MCI increment can be properly addressed by a modification of the 'isotropic' MCI theory relationships \cite{zhdanov2009}.  The results of such simple implementation are shown in Fig.~\ref{fig_3_spectra_theory}. Compared to the hexagonally symmetric HD distribution in the case of isotropic loading  [Fig.~\ref{fig_3_spectra_theory}(a)], the asymmetric loading [Figs.~\ref{fig_3_spectra_theory}(b) and~\ref{fig_3_spectra_theory}(c)] breaks the hexagonal symmetry.
For an appropriate orientation of the loading direction $\alpha$, the MCI is triggered in one direction, and only one pair of HD appears [see Fig.~\ref{fig_3_spectra_theory}(b)] and a single oscillation pattern dominates. If the MCI is triggered in two directions, there are two pairs of HD and, therefore, two competing synchronization patterns. These observations agree very well with the experiments and simulations \cite{couedel2014, laut2015}.

The distributions of Figs.~\ref{fig_3_spectra_theory}(a)--(c) explain fairly well all simulated and observed anisotropy effects but the rotational asymmetry of the measured spectra: The hot-dot 'twins' which are oriented in opposite directions one to another have exactly the same intensity in the model. This twofold symmetry is broken by adding a nonzero flux as we discuss below.

\subsection{Phonon flux} 
\label{subsec_flux}

The anisotropy of the compression phonon spectrum $I_{\mathbf{k},\omega}$ is directly related to the kinetic temperature gradient which is, in turn, proportional to the mean phonon flux $\langle\mathbf{u}\rangle$:
\begin{equation}\label{eq_Tgrad}
 \begin{gathered}
\left. \frac{\nabla T}{T} \right|_{\omega=\omega_\text{HD}}\propto\langle\mathbf{u}\rangle= 
\frac{\langle \mathbf{q} \rangle}{\mathcal{E}} \,, \\                                                          
\langle \mathbf{q} \rangle=\int_{\Delta\omega}d\omega \int dk_xdk_y\, \omega \, \mathbf{u}\, \mathcal{N}_{\mathbf{k},\omega},~~ 
\mathbf{u}=\partial_{\mathbf{k}}\omega \,, \\                                                
\mathcal{E}=\int_{\Delta\omega} d\omega \int dk_xdk_y\,\omega\, \mathcal{N}_{\mathbf{k},\omega} ,~~ 
\omega_{\mathbf{k}}\mathcal{N}_{\mathbf{k},\omega} = I_{\mathbf{k},\omega} \,.     
 \end{gathered}
\end{equation}
Here $\mathcal{N}_{\mathbf{k},\omega}$ is the phonon number density, $\langle \mathbf{q} \rangle$ is the mean energy flux, $\mathcal{E}$ the total wave energy, $\mathbf{u}$ the phonon speed, and $\Delta\omega$ the MCI-resonance width. (Here, and further on, $\hbar=1$ \cite{lifshitz1981}.)
The distances of the HDs inside the fBz from the origin are approximately the same, $k\simeq k_\text{HD}$, as well as the modulus of the phonon speeds $|\mathbf{u}|\simeq u_\text{HD}$. The phonon speed directions and the HD intensities are principally different, though, due to the anisotropy of the MCI of the deformed crystal \cite{laut2015}. It results in a nonzero energy flux, in distinction to the perfect crystals where the flux is zero by symmetry of the MCI \cite{zhdanov2009}. 

The main reason for a nonvanishing phonon flux in the distribution of $I_{\mathbf{k},\omega}$ shown in Fig.~\ref{fig_2_spectra_run1} is the energy difference of the quasisymmetric HDs, the 'twins'. There are only two such pairs of twins, one at $\theta \simeq 0^\circ$ and $\theta \simeq 180^\circ$, the other at $\theta\simeq 54^\circ$ and $\theta \simeq 234^\circ$. Using Eq.~\ref{eq_Tgrad}, the resulting phonon flux can be estimated from the spectrum shown in Fig.~\ref{fig_2_spectra_run1} as:
\begin{equation}
\frac{\langle\mathbf{u}\rangle}{u_\text{HD}}\approx-\left( 0.06 \mathbf{e}_1 + 0.13 \mathbf{e}_2 \right),
~~\frac{|\langle\mathbf{u}\rangle|}{u_\text{HD}}\approx 0.17
\end{equation}
where $\mathbf{e}_{1,2}=\frac{\mathbf{u}}{u}|_{\theta=0,\frac13\pi}$. The flux is normalized by the phonon speed $u_\text{HD}$ which remains unknown in this approach. It must be approximated differently, see below. The angle $\arg(\langle\mathbf{u}\rangle/u_\text{HD}) = 223^\circ$ is close to $\theta = 234^\circ$ as could be expected from Fig.~\ref{fig_2_spectra_run1}.

\begin{figure}
\includegraphics[width=.8\linewidth,trim = 6mm 6mm 9mm 9mm, clip]{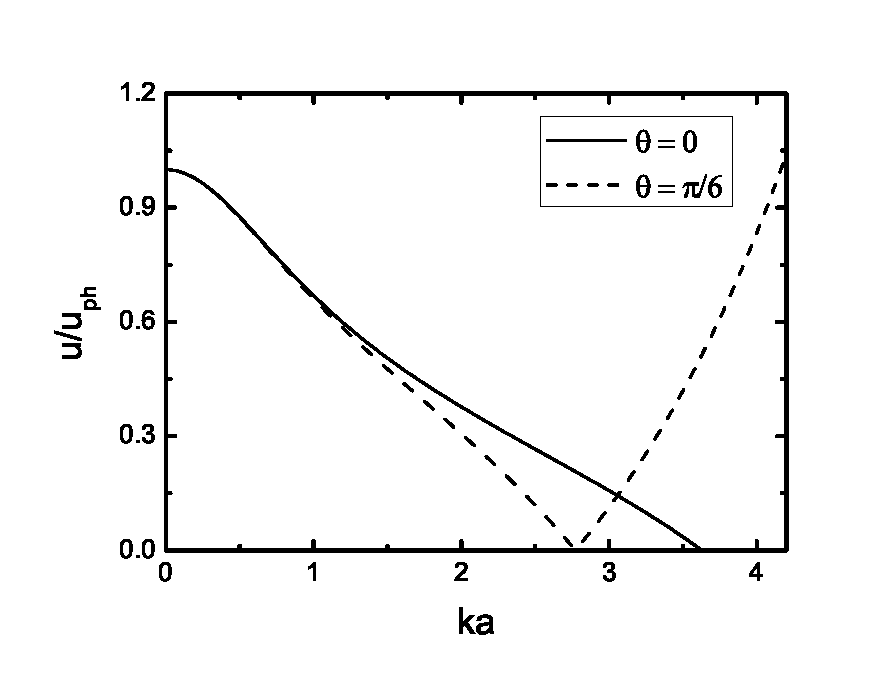} 
\caption{Compression wave group-to-phase velocity ratio $u/u_{ph}$ vs reduced wavenumber $ka$ calculated for the two directions $\theta = 0^\circ,\,60^\circ$ \cite{zhdanov2009}. For fig.~\ref{fig_2_spectra_run1}: $k_\text{HD}a=3.08$,  $u/u_{ph}|_{\theta=0}\simeq u/u_{ph}|_{\theta=\frac13\pi}\simeq14\%$.
} \label{fig_4_velocity_ratio}
\end{figure}

Adding a nonvanishing phonon drift to the theoretical model results in an asymmetric, direction-dependent spectrum, see Figs.~\ref{fig_3_spectra_theory}(d) and~\ref{fig_3_spectra_theory}(e). Qualitatively (detailed analysis will be published elsewhere) a weak nonzero drift, say, along the main instability direction would result in a difference of the maximal phonon energy of the order of $\delta\omega_h\approx 2k_\text{HD}u$, where $k_\text{HD}$ is the HD (i.e., resonant) wave number, $k_\text{HD}a<k_ba=2\pi/\sqrt{3}\simeq3.63$. Therefore, the resonant condition of the horizontal and vertical mode crossing would be satisfied a bit earlier at the 'hotter' edge of the fBz:
\begin{equation}\label{eq_SecondVariation}
\frac{\delta\Omega_{z, \text{crit}}}{\Omega_{z,\text{crit}}} \approx 
\frac{2k_\text{HD}u}{\omega_\text{HD}}=
2\xi_\text{HD}\frac{u}{u_\text{HD}} \,,
\end{equation}
where $\xi_\text{HD}=\left(u/u_{ph}\right)_\text{HD}$ is the compression wave group-to-phase velocity ratio (see Fig.~\ref{fig_4_velocity_ratio}). In the vicinity of the HDs, $u\ll \omega_\text{HD}/k_\text{HD}$ \cite{zhdanov2009}, therefore the magnitude of the effect is not large, as expected.

\section{Possible origin of symmetry breaking}
\label{sec_possible_origin_of_symmetry_braking}

The goal of this section is to properly address the question where the spectral asymmetry stems from. To answer this question it is necessary to thoroughly explore the main features of the HDs: (i) the structure of the velocity fluctuation spectra in reciprocal space, (ii) the energy distribution inside of the HDs, and (iii) the main dynamical processes responsible for the energy transport between the HDs.

\subsection{HD twins: the universality of the anisotropy mechanism} 

To analyze the spectral asymmetry, a second crystal with a more pronounced asymmetry is considered. The simulation Run~II was performed for a smaller number of particles $N=10000$, see Table~\ref{tab_1_params}. While the larger crystal of Run~I equilibrates to a structure with large defect lines around the center which reflect the sixfold symmetry of the lattice, the smaller crystal forms a less homogeneous dislocation pattern. The loading direction was set to $\alpha = 0^\circ$ which selects only the HDs along the $x$~axis at a weak MCI \cite{laut2015}. In order to activate the MCI also in the other directions, a smaller value of $f_z = 19.5$\,Hz was used during the Dynamical phase of Run~II.

\begin{figure}
\centering
\includegraphics[width=\columnwidth]{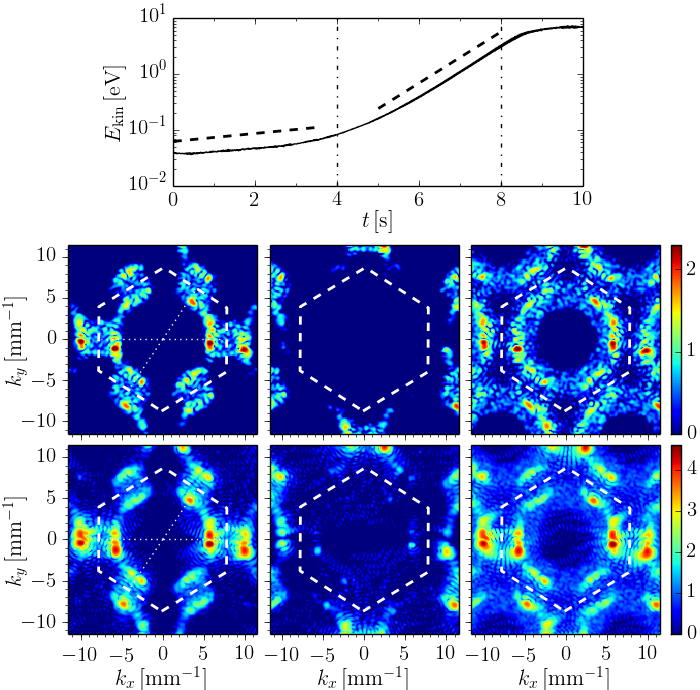}
\caption{
(Color online) Time evolution of velocity fluctuation spectra during the Dynamical phase of simulation Run~II (see Table~\ref{tab_1_params}).
Top panel: Kinetic energy as a function of time $t$ on a semi-logarithmic scale. The vertical dash-dotted lines at $t=4\,\mathrm{s}$ and $t=8\,\mathrm{s}$ indicate the time intervals used for the calculation of the spectra. The dashed lines correspond to growth rates $\dot E / E = 0.17\,\mathrm{s}^{-1}$ and $1.1\,\mathrm{s}^{-1}$.
Middle panels: Intensity of the fluctuation spectra $\bar I_\mathbf{k}$ for longitudinal (left), transverse horizontal (center) and transverse vertical (right) modes at $t=4\,\mathrm{s}$, averaged over the frequency range $15\,\mathrm{Hz} < f < 17\,\mathrm{Hz}$ and colorcoded on a logarithmic scale in arbitrary units. The white dashed line indicates the border of the first Brillouin zone. White dotted lines appear for the longitudinal mode at angles of $\theta = 0^\circ, \, 56^\circ$.
Bottom panels: The same for spectra calculated at $t=8\,\mathrm{s}$. Note a weak hot-dot-induced shear intensity in the transverse horizontal mode for $t=8$\,s. 
} \label{fig_5_spectra_run2}
\end{figure}

In the beginning of the Dynamical phase, the particle kinetic energy grows exponentially with a relatively small growth rate (see Fig~\ref{fig_5_spectra_run2}, top panel). The fluctuation energy of the monolayer starts to collapse, leading to the emergence of multiple HDs (see Fig.~\ref{fig_5_spectra_run2}, middle and bottom panels). After about 4\,s, the growth rate changes to a larger value. At $t\approx 8$\,s, the high kinetic energy of the particles leads to the breaking of the crystalline order.

\begin{figure}
\centering
\includegraphics[width=\columnwidth]{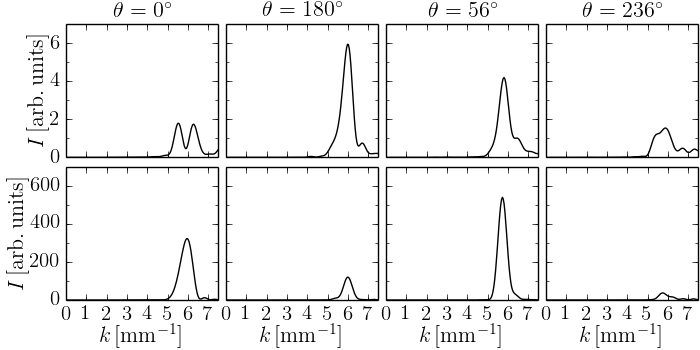}
\caption{
Intensities of fluctuation spectra in the main directions of the crystal for the simulated data of Run~II.
Top panels: $\bar I_{\mathbf{k}}$ calculated at $t=4\,\mathrm{s}$ at the angles of $\theta = 0^\circ,\, 180^\circ,\, 56^\circ,\, 236^\circ$, averaged over the frequency interval $15\,\mathrm{Hz} < f < 17\,\mathrm{Hz}$. Only the longitudinal mode was considered.
Bottom panels: The same for a later time step $t=8\,\mathrm{s}$.
} \label{fig_6_intensities_run2}
\end{figure}

The fluctuation spectrum reveals not only a pair of HD twins at $\theta\simeq0^\circ,\, 180^\circ$ as in Ref.~\cite{laut2015}, a comparatively weaker pair at  $\theta \simeq 56^\circ,\, 236^\circ$ is also present (see Fig~\ref{fig_5_spectra_run2}). The HD twins of the weaker pair have very different intensities. This feature, in particular, is useful to demonstrate the universality of the anisotropic MCI. Figure~\ref{fig_6_intensities_run2} shows the intensities of the HDs at two different time steps. The chirality of the fluctuation pattern becomes apparent when comparing the respective twins at $\theta = 0^\circ,\, 180^\circ$ and $\theta = 56^\circ,\, 236^\circ$.

\subsection{Hot dot energy distribution} 

\subsubsection{Hot dot core structure} 
Despite the asymmetry in the energy distribution between the HDs evidenced above, all HDs are equally, though quite delicately, structured. The frequency-averaged energy distribution $\bar I_k$ in the main directions of the crystal consists of a core and a turbulent halo, as can be seen in Fig~\ref{fig_7_hd_structure}. The core of the HD is well described by a Gaussian:
\begin{equation}\label{eq_gauss}
\bar I_k\propto \exp\left(-\frac{\left(k-k_{HD}\right)^2}{2\mu^2}\right).
\end{equation}
The core, by energy content, is the dominant part of the HD, and, therefore, the width of the Gaussian core $\mu=\langle\delta k_\text{HD}^2\rangle^\frac12$ can be ascribed to the size of the HD in $\mathbf{k}$ space. Typically, it is $\mu=0.2$--$0.4~\rm mm^{-1}$, that is, about 10--30 times smaller than the typical wave number of phonons comprising the HD, $\mu\ll k_\text{HD}\simeq 6~\rm mm^{-1}$; see Table~\ref{tab_3_hd_structure_params}. It is a crucial feature of the MCI in the weakly nonlinear regime. Such an islandlike distribution of the wave energy helps a lot to simplify the description of the wave dynamics. The gain of phonon energy is due to MCI while the loss is due to diffusion activated by phonon scattering \footnote{We simplified the description omitting the gradient term from the Fokker-Plank equation (\ref{eq_diff}). This term, though might be important to explain a weak asymmetry of the HD core (see, e.g., Fig.~\ref{fig_6_intensities_run2}), is small compared to the term accounting for diffusion.}:
\begin{equation}\label{eq_diff}
 \partial_t \bar I_k= \gamma_k^\text{MCI} \bar I_k + \mathcal{D}^{(k)} \partial^2_k \bar I_k \,,
\end{equation}
where $\gamma_k^\text{MCI}$ is the MCI increment and $\mathcal{D}^{(k)}$ is the diffusion coefficient in $\mathbf{k}$ space. Assuming a uniform energy gain, $\bar I_k \propto \exp(\gamma t)$, where $\gamma$ is the actual growth rate of fluctuations, and making use of relationship (\ref{eq_gauss}), it is easy to observe that 
\begin{equation}\label{eq_gamma}
 \gamma = \gamma_k^\text{MCI} + \frac{\mathcal{D}^{(k)}}{\mu^2} \left( \frac{\xi^2}{\mu^2} - 1\right),~ \xi= k - k_\text{HD} \,.
\end{equation}
The excitation region is limited in size,
\begin{equation}\label{eq_size}
 |k-k_\text{HD}|\leq
 \xi_{max}=\mu\sqrt{1+\frac{\gamma\mu^2}{\mathcal{D}^{(k)}}}\,,
\end{equation}
which is also in a fairly good agreement with theoretical model (see Fig.~\ref{fig_3_spectra_theory}).

To make a numerical example, let us consider the data from Run~II. The theory of Ref.~\cite{zhdanov2009} predicts $\max\left[\gamma_k^\text{MCI}\right]\simeq5.26~\rm s^{-1}$ for the parameter set of Run~II. The growth rate of the fluctuations can be approximated by the kinetic energy growth rate from Fig.~\ref{fig_5_spectra_run2}. Given the averaged HD size $\langle\mu\rangle=0.38~\rm mm^{-1}$ and $\gamma = 0.17\,\mathrm{s}^{-1}$ at $0\,\text{s}<t<4\,\text{s}$, from relationships (\ref{eq_gamma}) and (\ref{eq_size}) it follows immediately for the diffusion coefficient $\mathcal{D}^{(k)}\simeq 0.73~\rm mm^{-2} s^{-1}$ and for the size of the excitation region $\xi_{max}\simeq0.39~\rm mm^{-1}$. In the period $4\,\text{s}<t<8\,\text{s}$, given $\langle\mu\rangle=0.28~\rm mm^{-1}$ and $\gamma = 1.1\,\mathrm{s}^{-1}$, it yields a lower value $\mathcal{D}^{(k)}\simeq0.33~\rm mm^{-2} s^{-1}$, which is not surprising considering the enhanced energy growth rate. The size of the excitation region is estimated as  $\xi_{max}\simeq0.31~\rm mm^{-1}$. 

The growth rates $\gamma$ can also be obtained for each HD individually from the evolution of the fluctuation spectra, the resulting values for $\mathcal{D}^{(k)}$ and $\xi_{max}$ are shown in Table~\ref{tab_3_hd_structure_params}.

\subsubsection{HD turbulent halo} 
\begin{figure}
\centering
\includegraphics[width=0.78\linewidth, trim=4mm 4mm 6mm 8mm, clip]{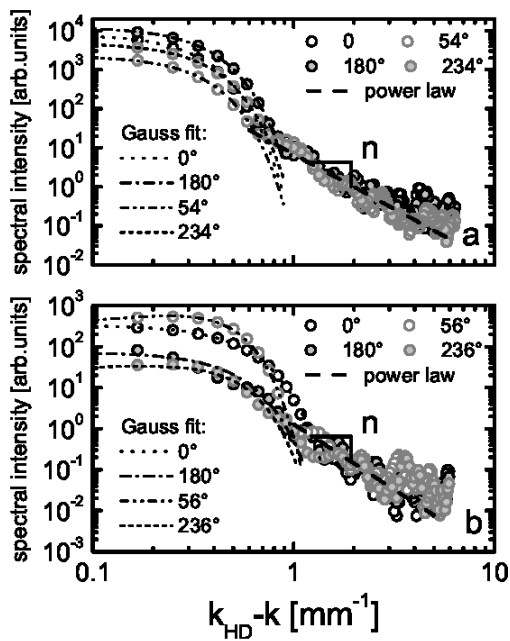} 
\caption{Fine structure of the hot-dot energy distribution $\bar I_k$ for different directions~$\theta$ as a function of deviation from the hot-dot center $k_\text{HD} - k$. 
(a) Run~I, $t=25$\,s of the Dynamical phase, see Fig.~\ref{fig_2_spectra_run1}. 
(b) Run~II, $t=8$\,s of the Dynamical phase, see Fig.~\ref{fig_6_intensities_run2}. The cores of the hot dots are individually fitted to Gauss distributions, the widths $\mu$ are collected in Table~\ref{tab_3_hd_structure_params}. The tails of the curves are fitted to a power law with exponent $n=2.80\pm0.08$ (Run~I) and $n=3.2\pm0.2$ (Run~II). 
} \label{fig_7_hd_structure}
\end{figure}

The turbulent suprathermal halo, essentially an isotropic feature associated with every HD, is well recognizable in the log-log plot of Fig.~\ref{fig_7_hd_structure} by an abrupt change in the slope of the energy spectrum. The fluctuation energy is power-law distributed in the halo, $\bar I_k\propto|k_\text{HD}-k|^{-n}$. It is worth noting that the exponent deviates not much from the value $n\simeq 3$ which is typical for frictional turbulence \cite{frisch1995, schwabe2014, zhdanov2015wave}. It is natural to associate the appearance of these quanta at least partly with the Umklapp scattering of high-energy HD phonons. For quasiequilibrium situations such kind of scattering process is well studied, see, e.g., Ref.~\cite{matveev2010}. Note that the halo intensifies with time at the nonlinear stage of MCI.

\begin{table} 
 \caption {
 Characteristics of the hot-dot energy distribution. The time-averaged growth rate $\gamma= \langle \dot I/I \rangle$ is calculated from the velocity fluctuation spectra (see Figs.~\ref{fig_2_spectra_run1} and \ref{fig_6_intensities_run2}) for the direction given in the second column. The hot-dot size in $k$~space, $\mu\equiv\langle\delta k_\text{HD}^2\rangle^{1/2}$, is obtained by using data of Fig.~\ref{fig_7_hd_structure}. The size of the excitation region $\xi_\mathrm{max}$ and the diffusion coefficient $\mathcal{D}^{(k)}$ in $k$ space are estimated with Eqs.~(\ref{eq_gamma}) and (\ref{eq_size}). $\gamma$ was calculated at $t=12.5$--$25$\,s (Run I) and $t=4$--$8$\,s (Run II). The relative errors are 10\% for $\gamma$, 5\% for $\mu$, 12\% for $\xi_\text{max}$ and 20\% for $\mathcal{D}^{(k)}$. 
 } \label{tab_3_hd_structure_params}
 \begin{ruledtabular}
  \begin{tabular}{l c  d                        c                      c                      c}
   Run & $\theta$   & \mc{$\gamma$}           & $\mu$                & $\xi_{max}$          & $\mathcal{D}^{(k)}$                \\
       & ($^\circ$) & \mc{($\mathrm{s}^{-1}$)}& ($\mathrm{mm}^{-1}$) & ($\mathrm{mm}^{-1}$) & ($\mathrm{mm}^{-2}\mathrm{s}^{-1}$)\\
   \colrule\noalign{\smallskip}
       & 0          & 0.34                    & 0.20                 & 0.21                 & 0.20                                      \\
   I   & 180        & 0.35                    & 0.22                 & 0.23                 & 0.24                                      \\
       & 54         & 0.33                    & 0.21                 & 0.22                 & 0.22                                      \\ 
       & 234        & 0.47                    & 0.22                 & 0.23                 & 0.23                                      \\
   \colrule\noalign{\smallskip}
       & 0          & 1.3                     & 0.35                 & 0.40                 & 0.49                                  \\
   II  & 180        & 0.75                    & 0.32                 & 0.35                 & 0.46                                \\
       & 56         & 1.2                     & 0.21                 & 0.24                 & 0.18                                 \\ 
       & 236        & 0.80                    & 0.26                 & 0.28                 & 0.30                                \\
  \end{tabular}
 \end{ruledtabular}
\end{table}

\subsection{Anisotropic phonon scattering by defects} 
\label{subsec_scattering_at_defects}

By virtue of relationship (\ref{eq_gamma}), since the actual growth rates $\gamma$ and the core sizes $\mu$ are only slightly different for HD twins (see Table~\ref{tab_3_hd_structure_params}), the intensity asymmetry might also stem from the anisotropic phonon diffusivity governed, e.g., by anisotropic interaction of phonons and dislocations, or by Umklapp processes that lead to a loss of high-energy HD quanta. We start with an analysis of the role of the defects. 

\subsubsection{Role of defects} 
Phonon scattering on defects, apart from nonlinear phonon interactions and finite-size effects, is known as one possible mechanism of energy redistribution between the phonons and their anisotropic transport \cite{yang2003, maurel2004, maurel2006}. It is also a well-known fact that the anisotropy of the thermal conductivity is closely connected to the special features of the phonon spectra determined by phonon scattering by oriented dislocations \cite{lugueva2004}. For instance, no interaction occurs between the longitudinal wave and the dislocation when an incident wave propagating in a direction parallel or perpendicular to the Burgers vector \cite{maurel2004}. 
If the phonon flux is normal to the orientation of chains of dislocations, the scattering is stronger \cite{maurel2006}. 

\begin{figure}
\includegraphics[width=\linewidth]{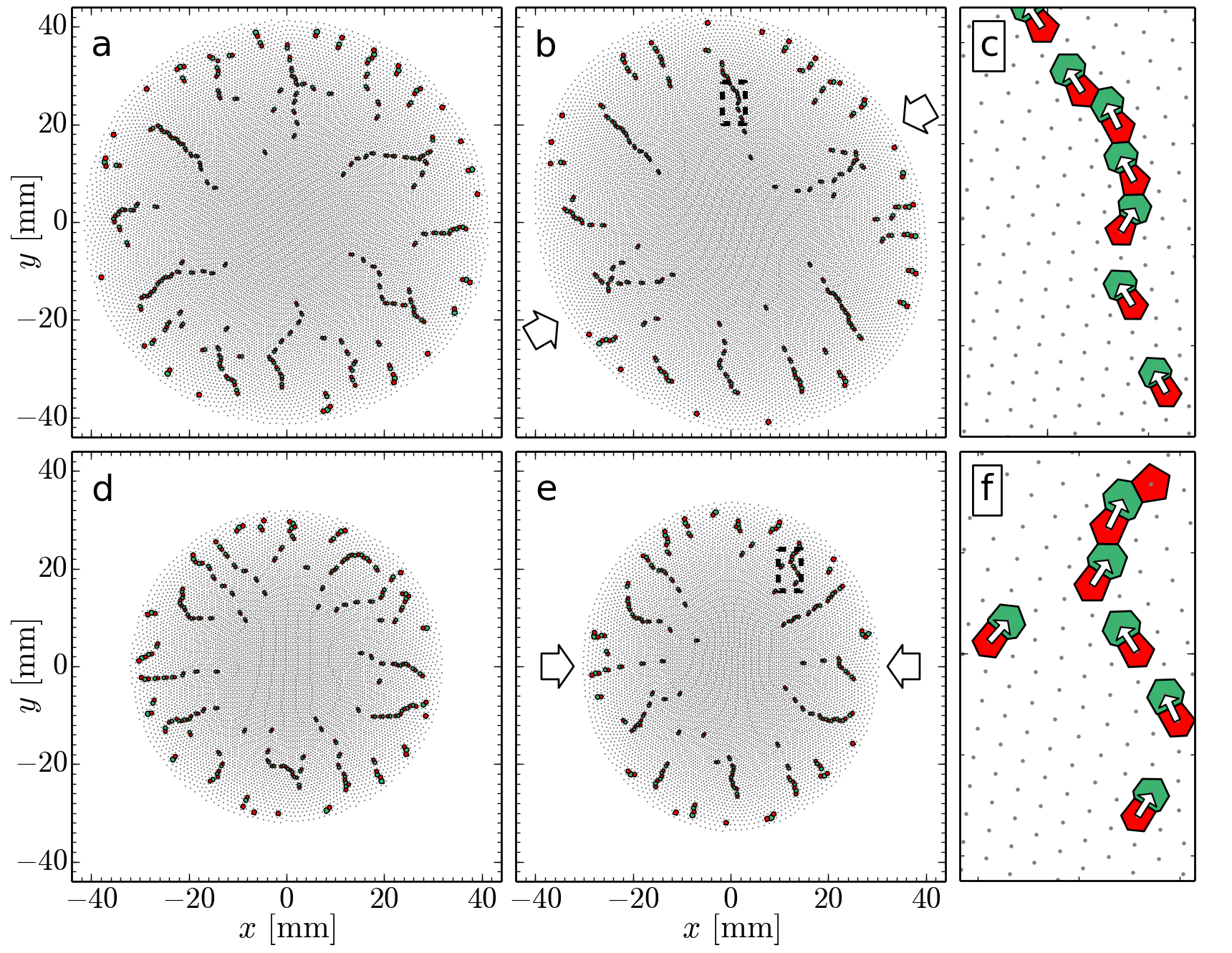}
\caption{
(Color online) Asymmetric dislocation pattern of the compressed crystals. 
Shown is the crystal of simulation Run I at the beginning of the Deformation phase (a) and at the beginning of the Dynamical phase (b). The dashed square is magnified by a factor of 10 in (c). The particle positions are shown as gray dots, and the Voronoi cells of the fivefold and sevenfold defects are shown in red (dark gray) and green (light gray), respectively. In (c), the dipoles $\mathbf{d}$ as defined in the text are also shown as white arrows. Note an apparent trend of dislocation chains to align transversally to the compression direction~$\alpha$, which is indicated by arrows in~(b). 
(d)--(f):~The dislocation patterns for Run~II.
} \label{fig_8_dislocation_pattern}
\end{figure}

The anisotropic heat transport in a plasma crystal has been studied in Refs.~\cite{nunomura2005heat, nosenko2008heat}. In our simulated crystals phonon scattering by defects might be quite well pronounced because the dislocation chains \cite{knapek2007, knapek2013} (or dislocation 'scars' \cite{ling2005}) that form during the Equilibration and Deformation phases exhibit a preferred orientation which tends to be perpendicular to the direction of dominant loading. This is evidenced in Fig.~\ref{fig_8_dislocation_pattern} where snapshots of the two simulation runs at the beginning of the Deformation phase and of the Dynamical phase are shown. 

If the dislocation pattern inside the crystal is random, the wave is only expected to be attenuated through diffusive scattering. If, however, the pattern is asymmetric, phonon scattering can lead to a broken parity symmetry. To measure the asymmetry in the dislocation pattern, we calculate the center of mass $\mathbf{R}_d = \langle \mathbf{r}_7 \rangle$, where $\mathbf{r}_7$ are the positions of the sevenfold defect cells. The apparent defects at the very border of the crystal are not considered for the calculation. Magnitude and argument of this vector are indicated in Table \ref{tab_4_dislocation_measures} for the defect patterns shown in Fig.~\ref{fig_8_dislocation_pattern}. It can be seen that both values change drastically during the Deformation phase of Run~I. Still, the magnitude $|\mathbf{R}_d|$ stays relatively small. Despite the smaller crystal size in Run~II, the value of $|\mathbf{R}_d|$ is larger, indicating a more pronounced inhomogeneity of the dislocation pattern.

\subsubsection{Polarized dislocation patterns} 
When considering the influence of defects, it is important to note that their positions in the monolayer follow certain patterns closely connected to the external confinement. Every dislocation consists of a coupled pair of sevenfold and fivefold cells. It is characterized by the Burgers vector $\mathbf{b}$, whose orientation defines the dislocation gliding direction \cite{kittel1976}, or, equivalently, by the dislocation dipole vector, traditionally introduced as $\mathbf{d}=\mathbf{r}_7-\mathbf{r}_5$ \cite{paulose2015}, where $\mathbf{r}_{7,5}$ are the positions of the centers of the sevenfold and fivefold defect cells. For a single dislocation in an otherwise ideal lattice, $\mathbf{d}=\mathbf{b}\times \mathbf{e}_z$. Any applied external force with a nonzero component along the Burgers vector (transversal to the dipole vector) causes dislocation \emph{glide} \cite{kittel1976}. (Dislocation transversal \emph{creep} is much less probable). For both Run~I and Run~II, the slip events are rare and the dislocation patterns are quasistationary even during the Dynamical phase. 

A curious peculiarity is evident at close observation of the dislocation pattern: The majority of the dislocation pairs are ordered in such a way that their fivefold components are located closer to the cluster center than their sevenfold counterparts [see Figs.~\ref{fig_8_dislocation_pattern}(c) and~\ref{fig_8_dislocation_pattern}(f)]. The system of dislocation dipoles is therefore polarized by the external confinement. To quantify this effect, the global polarization is calculated as $\mathbf{D} = \sum \mathbf{d}$, where the sum is performed over all polarization vectors $\mathbf{d}$. We follow a rather simple rule for counting the polarization vectors. In chains of more than two defects, going radially outward, each sevenfold defect is connected to at most one fivefold defect. In cases where there are more fivefold defects than sevenfold defects, as in the upper part of Fig.~\ref{fig_8_dislocation_pattern}(f), the outmost defect is thus not considered. The magnitude and argument of $\mathbf{D}$ are shown in Table~\ref{tab_4_dislocation_measures}. $|\mathbf{D}|$ much increases during the Deformation phase of Run~I. Similar to the development of $|\mathbf{R}_d|$, the magnitude of the average polarization slightly decreases from an initially relatively large value during the Deformation phase of Run~II. 

\begin{table}
 \caption{Center of mass $\mathbf{R}_d$ and polarization $\mathbf{D}$ of the dislocation patterns shown in Fig.~\ref{fig_8_dislocation_pattern}. See text for the definitions. 
 } \label{tab_4_dislocation_measures}
 \begin{ruledtabular} 
  \begin{tabular}{ l l        D{.}{.}{1.2}            c                       D{.}{.}{1.1}           c }
             &              & \mc{$|\mathbf{R}_d|$} & $\arg(\mathbf{R}_d)$  & \mc{$|\mathbf{D}|$}  & $\arg(\mathbf{D})$ \\
   Run       & Phase        & \mc{(mm)}             & ($^\circ$)            & \mc{(mm)}            & ($^\circ$)  \\
   \colrule\noalign{\smallskip}
   I         & Deformation  & 0.57                  & 110                   & 1.0                  & 354       \\
   I         & Dynamical    & 1.3                   & 19                    & 5.6                  & 14         \\
   II        & Deformation  & 1.3                   & 81                    & 5.7                  & 76        \\
   II        & Dynamical    & 1.1                   & 90                    & 4.3                  & 169       \\
  \end{tabular}
 \end{ruledtabular}
\end{table}

\subsection{Asymmetric Umklapp scattering} 
\label{subsec_asymmetric_u_scattering}

If the crystal is perfect, without defects or strains in its structure, the scattering of phonons will only be caused by three-phonon processes in which two phonons coalesce to give one, or one splits up to give two \cite{peierls1929, berman1951}. Such nonlinear phonon scattering can be described by integral equations  (see, e.g., \cite{lifshitz1981}) which take into account the \emph{Umklapp} processes, or U-processes, that result from the periodicity of the lattice \cite{berman1951, peierls1929}. Three-phonon U-processes are known as the main intrinsic thermalresistive processes in crystals \cite{han1996}. They are apparently important and unavoidable during MCI in a plasma crystal because this instability generates phonons dominantly in the very proximity of the fBz boundary \cite{zhdanov2009}. For instance, in Fig~\ref{fig_2_spectra_run1}, the energy of the wave fluctuations is concentrated at $k_\text{HD}a>k_b a/2$, which is close to the fBz boundary $k_ba = 2\pi/\sqrt{3}$. The second harmonic must be expected at $k_sa \simeq 2k_\text{HD}a-2k_ba<0$, that is, in the opposite direction due to a U-process. The high-energy fundamental phonons, when coalesce, should formally disappear. On the other hand, elimination of high-energy phonons by Umklapp processes can be well compensated by the generation caused by the MCI. 

The U-processes are less important at the initial stage of the instability since U-processes are three-wave interactions leading to the generation of second harmonics, hence, they are at least of the second order by perturbation amplitude \cite{bloembergen1965}. At this stage, the anisotropic scattering by dislocations could be the only cause of spectral asymmetry. 

Further on, at the weakly nonlinear stage of MCI, U-scattering intensifies and, in analogy with \cite{matveev2010}, scattering of HD phonons caused by U-processes leads mainly to the relaxation of their distribution function, that is, to an effective nonlinear damping. The phonon fluxes of two hot-dot twins are not negligible, though mainly counter-directed. An important contribution to the phonon drift is thus only possible if the spectral pattern is asymmetric.

\section{conclusion}
\label{sec_conclusion}

The main features in the particle current fluctuation spectra (which are also observed in experiments \cite{couedel2014}) were reproduced by a simple theoretical model incorporating the angle-dependence of the crystal interaction range. 
An anisotropic confinement of the crystal enhances the MCI increment in the direction of the compression and leads to hot dots of different intensities. The broken left-right symmetry of pairs of hot dots (twins) was reproduced by taking into account a nonvanishing phonon flux.

Two possible explanations for such a mean phonon flux were presented. The first one is the Umklapp process resulting in a turbulent power-law distributed halo surrounding the hot dots. Phonon scattering by defects was presented as a second mechanism producing a phonon flux. The analysis of the dislocation pattern showed that most pairs of fivefold and sevenfold defects are 'polarized' such that their fivefold components are located closer to the cluster center. Simple measures relying on the center of mass and dipole moment of the defect pattern were proposed to quantify the influence of this effect. A more detailed analysis of the structure of the defect chains will be necessary for further insights.

It depends mostly on the symmetry of the hot-dot positions in the first Brillouin zone whether the total phonon drift caused by Umklapp processes cancels or not. For a highly ordered hot-dot pattern (see Fig.~\ref{fig_2_spectra_run1}) this is certainly the case, and the nonvanishing phonon drift originates mainly from an anisotropic scattering by dislocations rather than Umklapp processes.

For a less symmetric pattern of hot dots (see Fig.~\ref{fig_5_spectra_run2}), the situation is not that simple. The hot-dot twins are not only different in their energy content, but they are also positioned asymmetrically in the first Brillouin zone. Umklapp scattering may thus be another reason for the systematic phonon drift in Run~II. The different growth rates of the kinetic energy observed in Fig.~\ref{fig_5_spectra_run2} may hint towards a transition from an initial regime where the phonon flux is dominated by scattering on defects to a regime where three-body Umklapp scattering plays an important role. Analyzing higher harmonics of the hot dots may give additional insights, it will be subject to further studies. 

To conclude, we have analyzed the spectral asymmetry of compressed plasma crystals. The finiteness of the crystal was explicitly taken into account, since it has an impact on the hot-dot positions in reciprocal space and enables the formation of an ordered dislocation pattern. Both effects can explain the spectral asymmetry observed in experiments and simulations.

\begin{acknowledgments}
We wish to thank L\'ena\"ic Cou\"edel for helpful discussions.
This work was supported by 
the German Federal Ministry for Economy and Technology under grant No.~50WM1441
and by
the European Research Council under the European Union's Seventh Framework Programme (FP7/2007-2013) / ERC Grant agreement 267499.
\end{acknowledgments}

\begin{appendix}

\section{Interaction energy minimization}
\label{app1}
The energy of the two dimensional $N$-particle 'cold' cluster (assuming the infinitely strong vertical confinement) in a parabolic well is
\begin{equation}\label{eq_Energy}
W=\frac12Q^2\sum^N_{i\neq j}R_{i,j}^{-1}\exp\left(-\lambda^{-1}R_{i,j}\right)+\frac12M\Omega_c^2\sum^N_ir_i^2,
\end{equation}
where $R_{i,j}=|\mathbf{r}_i-\mathbf{r}_j|$, $i,j=1\ldots N$, $N$ the number of particles. To minimize the cluster energy, one has to consider the system of N equations $\delta_{\mathbf{r}_i} W=0,~i=1\ldots N$, which can be solved numerically provided that N is not too large. Equation (\ref{eq_Energy}) can be significantly simplified in the mean-field approximation. The sums over particle positions are replaced by integrals over the particle number density per unit area, $n(\mathbf{r})$, where the integration is performed over the cluster area $S_c$ \cite{totsuji2001}:
\begin{equation}\label{eq_ContEnergy}
 \begin{gathered}
W=W_{int}+W_{ext}, \\
W_{int}=\frac12Q^2\int_{S_c} d\mathbf{r}'d\mathbf{r}n(\mathbf{r}')n(\mathbf{r})R^{-1}\exp\left(-\lambda^{-1}R\right), \\
W_{ext}=\frac12 M\Omega_c^2\int_{S_c} d\mathbf{r}n(\mathbf{r})r^2, \\
N=\int_{S_c} d\mathbf{r}n(\mathbf{r}).
 \end{gathered}
\end{equation}
In order to validate relationship (\ref{eq_kappa}), it is enough to consider a uniform number density distribution,
\begin{equation}\label{eq_Uniform}
n(r) = \left\{ \begin{array}{rcl}
\langle n\rangle & \mbox{for}
& r\leq R_c \\ 0 & \mbox{for} & r>R_c
\end{array}\right.,
\end{equation}
where $R_c$ is the cluster radius and $\langle n\rangle\propto a^{-2}$ the mean cluster number density. Under this assumption, one obtains
\begin{equation}\label{eq_UniformEnergy}
N=\langle n\rangle S_c,~
W_{ext}=\frac{1}{4\pi}MNS_c\Omega_c^2.
\end{equation}
To calculate $W_{int}$, let us recall the useful relationship:
\begin{equation}\label{eq_Bessel}
 \begin{gathered}
R^{-1}\exp\left(-\lambda^{-1}R\right)=\lambda\int_{0}^{\infty}\frac{kdk}{\sqrt{1+k^2\lambda^2}}
J_0\left(kR\right), \\
J_0\left(kR\right)=J_0\left(kr\right)J_0\left(kr'\right)+2\sum_{s=1}^\infty J_s\left(kr\right)J_s\left(kr'\right)\cos(\phi), \\
R=\sqrt{r^2+r'^2-2rr'\cos(\phi)},~~
 \end{gathered}
\end{equation}
where $J_s$ is the Bessel function. Then
\begin{equation}\label{eq_Wint}
W_{int}=\frac{q^2N^2}{\lambda}F(\xi),\, F(\xi)=\int_0^\infty\frac{2\xi dk}{k\sqrt{1+\xi k^2}}J_1^2(k),
\end{equation}
where $\xi=S_\lambda/S_c$, $S_\lambda=\pi\lambda^2$. For large cluster (as normally the case in experiments and simulations) $\xi\ll 1$, therefore $F(\xi)\simeq \xi$, $W_{int}\simeq\frac{q^2N^2}{\lambda}\xi$, and the total cluster energy is
\begin{equation}\label{eq_TotalEnergy}
W=W_{ext}+W_{int}=\frac{1}{4\pi}MNS_c\Omega_c^2+\frac{q^2N^2}{\lambda}\frac{S_\lambda}{S_c}.
\end{equation}
The total energy $W$ as a function of $S_c$ has a minimum at
\begin{equation}\label{eq_Min}
S_c=2S_\lambda\sqrt{N}\frac{\Omega_q}{\Omega_c},
\end{equation}
if all other parameters are kept fixed. Since $S_c/S_\lambda\propto\kappa^2 $ we have $\kappa^2\propto\Omega_c^{-1}$, restoring Eq.~(\ref{eq_kappa}).

\section{Squeezed cluster: Eccentricity of the structure}
\label{app2}
Let us consider the slightly deformed crystal assuming an elliptic-shaped confining well:
\begin{equation}\label{eq_OmegaC}
\Omega_c=\Omega_c(\theta)=\Omega_{c,0}\sqrt{1+p\cos(2\theta)}, ~~p\ll 1,
\end{equation}
with $p$ as an asymmetry measure and an eccentricity:
\begin{equation}\label{eq_eC}
e^2_c=\frac{2p}{1+p}\equiv 1-\frac{\Omega^2_{c,min}}{\Omega^2_{c,max}},
\end{equation}
By virtue of Eq.~(\ref{eq_kappa}), the crystal interaction range is also weakly angle-dependent:
\begin{equation}\label{eq_kappa1}
\kappa=\kappa(\theta)=\frac{\kappa_{0}}{\sqrt[4]{1+p\cos(2\theta)}}.
\end{equation}
The eccentricity of this distribution is
\begin{equation}\label{eq_eKappa}
e^2_\kappa= 1-\frac{\kappa^2_{c,min}}{\kappa^2_{c,max}}\equiv1-\frac{\Omega_{c,min}}{\Omega_{c,max}}.
\end{equation}
In Eqs.~(\ref{eq_OmegaC}) and (\ref{eq_kappa}), $\Omega_{c,0}$ and $\kappa_{0}$ are angle-independent constants. They can be related to the 'unperturbed' crystal. For instance, for a pure shear deformation a constraint
\begin{equation}\label{eq_newlabel}
2\pi \kappa^2_{p=0}=\int^{2\pi}_0\kappa^2(\theta)d\theta=\kappa_0^2 \int^{2\pi}_0\frac{d\theta}{\sqrt{1+p\cos(2\theta)}}.
\end{equation}
allows one to obtain the interaction range $\kappa_0=\kappa_0(p)$ for any given asymmetry parameter $p$ through $\kappa^2_{p=0}$ of the unperturbed crystal.

\end{appendix}

\bibliography{./../literature}

\begin{thebibliography}{61}
\expandafter\ifx\csname natexlab\endcsname\relax\def\natexlab#1{#1}\fi
\expandafter\ifx\csname bibnamefont\endcsname\relax
  \def\bibnamefont#1{#1}\fi
\expandafter\ifx\csname bibfnamefont\endcsname\relax
  \def\bibfnamefont#1{#1}\fi
\expandafter\ifx\csname citenamefont\endcsname\relax
  \def\citenamefont#1{#1}\fi
\expandafter\ifx\csname url\endcsname\relax
  \def\url#1{\texttt{#1}}\fi
\expandafter\ifx\csname urlprefix\endcsname\relax\def\urlprefix{URL }\fi
\providecommand{\bibinfo}[2]{#2}
\providecommand{\eprint}[2][]{\url{#2}}

\bibitem[{\citenamefont{Chu and I}(1994)}]{chu1994}
\bibinfo{author}{\bibfnamefont{J.~H.} \bibnamefont{Chu}} \bibnamefont{and}
  \bibinfo{author}{\bibfnamefont{L.}~\bibnamefont{I}}, \bibinfo{journal}{Phys.
  Rev. Lett.} \textbf{\bibinfo{volume}{72}}, \bibinfo{pages}{4009}
  (\bibinfo{year}{1994}).

\bibitem[{\citenamefont{Thomas et~al.}(1994)\citenamefont{Thomas, Morfill,
  Demmel, Goree, Feuerbacher, and M{\"o}hlmann}}]{thomas1994}
\bibinfo{author}{\bibfnamefont{H.}~\bibnamefont{Thomas}},
  \bibinfo{author}{\bibfnamefont{G.~E.} \bibnamefont{Morfill}},
  \bibinfo{author}{\bibfnamefont{V.}~\bibnamefont{Demmel}},
  \bibinfo{author}{\bibfnamefont{J.}~\bibnamefont{Goree}},
  \bibinfo{author}{\bibfnamefont{B.}~\bibnamefont{Feuerbacher}},
  \bibnamefont{and}
  \bibinfo{author}{\bibfnamefont{D.}~\bibnamefont{M{\"o}hlmann}},
  \bibinfo{journal}{Phys. Rev. Lett.} \textbf{\bibinfo{volume}{73}},
  \bibinfo{pages}{652} (\bibinfo{year}{1994}).

\bibitem[{\citenamefont{Hayashi and Tachibana}(1994)}]{hayashi1994}
\bibinfo{author}{\bibfnamefont{Y.}~\bibnamefont{Hayashi}} \bibnamefont{and}
  \bibinfo{author}{\bibfnamefont{K.}~\bibnamefont{Tachibana}},
  \bibinfo{journal}{Jpn. J. Appl. Phys.} \textbf{\bibinfo{volume}{33}},
  \bibinfo{pages}{L804} (\bibinfo{year}{1994}).

\bibitem[{\citenamefont{Fortov et~al.}(2005)\citenamefont{Fortov, Ivlev,
  Khrapak, Khrapak, and Morfill}}]{fortov2005}
\bibinfo{author}{\bibfnamefont{V.~E.} \bibnamefont{Fortov}},
  \bibinfo{author}{\bibfnamefont{A.~V.} \bibnamefont{Ivlev}},
  \bibinfo{author}{\bibfnamefont{S.~A.} \bibnamefont{Khrapak}},
  \bibinfo{author}{\bibfnamefont{A.~G.} \bibnamefont{Khrapak}},
  \bibnamefont{and} \bibinfo{author}{\bibfnamefont{G.~E.}
  \bibnamefont{Morfill}}, \bibinfo{journal}{Phys. Rep.}
  \textbf{\bibinfo{volume}{421}}, \bibinfo{pages}{1} (\bibinfo{year}{2005}).

\bibitem[{\citenamefont{Morfill and Ivlev}(2009)}]{morfill2009}
\bibinfo{author}{\bibfnamefont{G.~E.} \bibnamefont{Morfill}} \bibnamefont{and}
  \bibinfo{author}{\bibfnamefont{A.~V.} \bibnamefont{Ivlev}},
  \bibinfo{journal}{Rev. Mod. Phys.} \textbf{\bibinfo{volume}{81}},
  \bibinfo{pages}{1353} (\bibinfo{year}{2009}).

\bibitem[{\citenamefont{Thomas and Morfill}(1996)}]{thomas1996}
\bibinfo{author}{\bibfnamefont{H.~M.} \bibnamefont{Thomas}} \bibnamefont{and}
  \bibinfo{author}{\bibfnamefont{G.~E.} \bibnamefont{Morfill}},
  \bibinfo{journal}{Nature (London)} \textbf{\bibinfo{volume}{379}},
  \bibinfo{pages}{806} (\bibinfo{year}{1996}).

\bibitem[{\citenamefont{Schweigert et~al.}(1998)\citenamefont{Schweigert,
  Schweigert, Melzer, Homann, and Piel}}]{schweigert1998}
\bibinfo{author}{\bibfnamefont{V.~A.} \bibnamefont{Schweigert}},
  \bibinfo{author}{\bibfnamefont{I.~V.} \bibnamefont{Schweigert}},
  \bibinfo{author}{\bibfnamefont{A.}~\bibnamefont{Melzer}},
  \bibinfo{author}{\bibfnamefont{A.}~\bibnamefont{Homann}}, \bibnamefont{and}
  \bibinfo{author}{\bibfnamefont{A.}~\bibnamefont{Piel}},
  \bibinfo{journal}{Phys. Rev. Lett.} \textbf{\bibinfo{volume}{80}},
  \bibinfo{pages}{5345} (\bibinfo{year}{1998}).

\bibitem[{\citenamefont{Nunomura et~al.}(2000)\citenamefont{Nunomura, Samsonov,
  and Goree}}]{nunomura2000}
\bibinfo{author}{\bibfnamefont{S.}~\bibnamefont{Nunomura}},
  \bibinfo{author}{\bibfnamefont{D.}~\bibnamefont{Samsonov}}, \bibnamefont{and}
  \bibinfo{author}{\bibfnamefont{J.}~\bibnamefont{Goree}},
  \bibinfo{journal}{Phys. Rev. Lett.} \textbf{\bibinfo{volume}{84}},
  \bibinfo{pages}{5141} (\bibinfo{year}{2000}).

\bibitem[{\citenamefont{Misawa et~al.}(2001)\citenamefont{Misawa, Ohno, Asano,
  Sawai, Takamura, and Kaw}}]{misawa2001}
\bibinfo{author}{\bibfnamefont{T.}~\bibnamefont{Misawa}},
  \bibinfo{author}{\bibfnamefont{N.}~\bibnamefont{Ohno}},
  \bibinfo{author}{\bibfnamefont{K.}~\bibnamefont{Asano}},
  \bibinfo{author}{\bibfnamefont{M.}~\bibnamefont{Sawai}},
  \bibinfo{author}{\bibfnamefont{S.}~\bibnamefont{Takamura}}, \bibnamefont{and}
  \bibinfo{author}{\bibfnamefont{P.~K.} \bibnamefont{Kaw}},
  \bibinfo{journal}{Phys. Rev. Lett.} \textbf{\bibinfo{volume}{86}},
  \bibinfo{pages}{1219} (\bibinfo{year}{2001}).

\bibitem[{\citenamefont{Nunomura et~al.}(2005)\citenamefont{Nunomura, Samsonov,
  Zhdanov, and Morfill}}]{nunomura2005heat}
\bibinfo{author}{\bibfnamefont{S.}~\bibnamefont{Nunomura}},
  \bibinfo{author}{\bibfnamefont{D.}~\bibnamefont{Samsonov}},
  \bibinfo{author}{\bibfnamefont{S.}~\bibnamefont{Zhdanov}}, \bibnamefont{and}
  \bibinfo{author}{\bibfnamefont{G.}~\bibnamefont{Morfill}},
  \bibinfo{journal}{Phys. Rev. Lett.} \textbf{\bibinfo{volume}{95}},
  \bibinfo{pages}{025003} (\bibinfo{year}{2005}).

\bibitem[{\citenamefont{Nosenko et~al.}(2008)\citenamefont{Nosenko, Zhdanov,
  Ivlev, Morfill, Goree, and Piel}}]{nosenko2008heat}
\bibinfo{author}{\bibfnamefont{V.}~\bibnamefont{Nosenko}},
  \bibinfo{author}{\bibfnamefont{S.}~\bibnamefont{Zhdanov}},
  \bibinfo{author}{\bibfnamefont{A.~V.} \bibnamefont{Ivlev}},
  \bibinfo{author}{\bibfnamefont{G.}~\bibnamefont{Morfill}},
  \bibinfo{author}{\bibfnamefont{J.}~\bibnamefont{Goree}}, \bibnamefont{and}
  \bibinfo{author}{\bibfnamefont{A.}~\bibnamefont{Piel}},
  \bibinfo{journal}{Phys. Rev. Lett.} \textbf{\bibinfo{volume}{100}},
  \bibinfo{pages}{025003} (\bibinfo{year}{2008}).

\bibitem[{\citenamefont{Menzel et~al.}(2010)\citenamefont{Menzel, Arp, and
  Piel}}]{menzel2010}
\bibinfo{author}{\bibfnamefont{K.~O.} \bibnamefont{Menzel}},
  \bibinfo{author}{\bibfnamefont{O.}~\bibnamefont{Arp}}, \bibnamefont{and}
  \bibinfo{author}{\bibfnamefont{A.}~\bibnamefont{Piel}},
  \bibinfo{journal}{Phys. Rev. Lett.} \textbf{\bibinfo{volume}{104}},
  \bibinfo{pages}{235002} (\bibinfo{year}{2010}).

\bibitem[{\citenamefont{Williams}(2014)}]{williams2014}
\bibinfo{author}{\bibfnamefont{J.~D.} \bibnamefont{Williams}},
  \bibinfo{journal}{Phys. Rev. E} \textbf{\bibinfo{volume}{90}},
  \bibinfo{pages}{043103} (\bibinfo{year}{2014}).

\bibitem[{\citenamefont{Ikezi}(1986)}]{ikezi1986}
\bibinfo{author}{\bibfnamefont{H.}~\bibnamefont{Ikezi}},
  \bibinfo{journal}{Phys. Fluids} \textbf{\bibinfo{volume}{29}},
  \bibinfo{pages}{1764} (\bibinfo{year}{1986}).

\bibitem[{\citenamefont{Melzer et~al.}(1999)\citenamefont{Melzer, Schweigert,
  and Piel}}]{melzer1999}
\bibinfo{author}{\bibfnamefont{A.}~\bibnamefont{Melzer}},
  \bibinfo{author}{\bibfnamefont{V.~A.} \bibnamefont{Schweigert}},
  \bibnamefont{and} \bibinfo{author}{\bibfnamefont{A.}~\bibnamefont{Piel}},
  \bibinfo{journal}{Phys. Rev. Lett.} \textbf{\bibinfo{volume}{83}},
  \bibinfo{pages}{3194} (\bibinfo{year}{1999}).

\bibitem[{\citenamefont{Ivlev and Morfill}(2000)}]{ivlev2000}
\bibinfo{author}{\bibfnamefont{A.~V.} \bibnamefont{Ivlev}} \bibnamefont{and}
  \bibinfo{author}{\bibfnamefont{G.}~\bibnamefont{Morfill}},
  \bibinfo{journal}{Phys. Rev. E} \textbf{\bibinfo{volume}{63}},
  \bibinfo{pages}{016409} (\bibinfo{year}{2000}).

\bibitem[{\citenamefont{Zhdanov et~al.}(2009)\citenamefont{Zhdanov, Ivlev, and
  Morfill}}]{zhdanov2009}
\bibinfo{author}{\bibfnamefont{S.~K.} \bibnamefont{Zhdanov}},
  \bibinfo{author}{\bibfnamefont{A.~V.} \bibnamefont{Ivlev}}, \bibnamefont{and}
  \bibinfo{author}{\bibfnamefont{G.}~\bibnamefont{Morfill}},
  \bibinfo{journal}{Phys. Plasmas} \textbf{\bibinfo{volume}{16}},
  \bibinfo{pages}{083706} (\bibinfo{year}{2009}).

\bibitem[{\citenamefont{Cou{\"e}del et~al.}(2010)\citenamefont{Cou{\"e}del,
  Nosenko, Ivlev, Zhdanov, Thomas, and Morfill}}]{couedel2010}
\bibinfo{author}{\bibfnamefont{L.}~\bibnamefont{Cou{\"e}del}},
  \bibinfo{author}{\bibfnamefont{V.}~\bibnamefont{Nosenko}},
  \bibinfo{author}{\bibfnamefont{A.~V.} \bibnamefont{Ivlev}},
  \bibinfo{author}{\bibfnamefont{S.~K.} \bibnamefont{Zhdanov}},
  \bibinfo{author}{\bibfnamefont{H.~M.} \bibnamefont{Thomas}},
  \bibnamefont{and} \bibinfo{author}{\bibfnamefont{G.~E.}
  \bibnamefont{Morfill}}, \bibinfo{journal}{Phys. Rev. Lett.}
  \textbf{\bibinfo{volume}{104}}, \bibinfo{pages}{195001}
  (\bibinfo{year}{2010}).

\bibitem[{\citenamefont{Cou{\"e}del et~al.}(2011)\citenamefont{Cou{\"e}del,
  Zhdanov, Ivlev, Nosenko, Thomas, and Morfill}}]{couedel2011}
\bibinfo{author}{\bibfnamefont{L.}~\bibnamefont{Cou{\"e}del}},
  \bibinfo{author}{\bibfnamefont{S.~K.} \bibnamefont{Zhdanov}},
  \bibinfo{author}{\bibfnamefont{A.~V.} \bibnamefont{Ivlev}},
  \bibinfo{author}{\bibfnamefont{V.}~\bibnamefont{Nosenko}},
  \bibinfo{author}{\bibfnamefont{H.~M.} \bibnamefont{Thomas}},
  \bibnamefont{and} \bibinfo{author}{\bibfnamefont{G.~E.}
  \bibnamefont{Morfill}}, \bibinfo{journal}{Phys. Plasmas}
  \textbf{\bibinfo{volume}{18}}, \bibinfo{pages}{083707}
  (\bibinfo{year}{2011}).

\bibitem[{\citenamefont{Cou{\"e}del et~al.}(2014)\citenamefont{Cou{\"e}del,
  Zhdanov, Nosenko, Ivlev, Thomas, and Morfill}}]{couedel2014}
\bibinfo{author}{\bibfnamefont{L.}~\bibnamefont{Cou{\"e}del}},
  \bibinfo{author}{\bibfnamefont{S.}~\bibnamefont{Zhdanov}},
  \bibinfo{author}{\bibfnamefont{V.}~\bibnamefont{Nosenko}},
  \bibinfo{author}{\bibfnamefont{A.~V.} \bibnamefont{Ivlev}},
  \bibinfo{author}{\bibfnamefont{H.~M.} \bibnamefont{Thomas}},
  \bibnamefont{and} \bibinfo{author}{\bibfnamefont{G.~E.}
  \bibnamefont{Morfill}}, \bibinfo{journal}{Phys. Rev. E}
  \textbf{\bibinfo{volume}{89}}, \bibinfo{pages}{053108}
  (\bibinfo{year}{2014}).

\bibitem[{\citenamefont{Reichhardt and Reichhardt}(2004)}]{reichhardt2004}
\bibinfo{author}{\bibfnamefont{C.}~\bibnamefont{Reichhardt}} \bibnamefont{and}
  \bibinfo{author}{\bibfnamefont{C.~J.~O.} \bibnamefont{Reichhardt}},
  \bibinfo{journal}{EPL} \textbf{\bibinfo{volume}{68}}, \bibinfo{pages}{303}
  (\bibinfo{year}{2004}).

\bibitem[{\citenamefont{Bohlein and Bechinger}(2012)}]{bohlein2012}
\bibinfo{author}{\bibfnamefont{T.}~\bibnamefont{Bohlein}} \bibnamefont{and}
  \bibinfo{author}{\bibfnamefont{C.}~\bibnamefont{Bechinger}},
  \bibinfo{journal}{Phys. Rev. Lett.} \textbf{\bibinfo{volume}{109}},
  \bibinfo{pages}{058301} (\bibinfo{year}{2012}).

\bibitem[{\citenamefont{Laut et~al.}(2015)\citenamefont{Laut, R\"ath, Zhdanov,
  Nosenko, Cou\"edel, and Thomas}}]{laut2015}
\bibinfo{author}{\bibfnamefont{I.}~\bibnamefont{Laut}},
  \bibinfo{author}{\bibfnamefont{C.}~\bibnamefont{R\"ath}},
  \bibinfo{author}{\bibfnamefont{S.}~\bibnamefont{Zhdanov}},
  \bibinfo{author}{\bibfnamefont{V.}~\bibnamefont{Nosenko}},
  \bibinfo{author}{\bibfnamefont{L.}~\bibnamefont{Cou\"edel}},
  \bibnamefont{and} \bibinfo{author}{\bibfnamefont{H.~M.}
  \bibnamefont{Thomas}}, \bibinfo{journal}{EPL} \textbf{\bibinfo{volume}{110}},
  \bibinfo{pages}{65001} (\bibinfo{year}{2015}).

\bibitem[{\citenamefont{Samsonov et~al.}(2005)\citenamefont{Samsonov, Zhdanov,
  and Morfill}}]{samsonov2005}
\bibinfo{author}{\bibfnamefont{D.}~\bibnamefont{Samsonov}},
  \bibinfo{author}{\bibfnamefont{S.}~\bibnamefont{Zhdanov}}, \bibnamefont{and}
  \bibinfo{author}{\bibfnamefont{G.}~\bibnamefont{Morfill}},
  \bibinfo{journal}{Phys. Rev. E} \textbf{\bibinfo{volume}{71}},
  \bibinfo{pages}{026410} (\bibinfo{year}{2005}).

\bibitem[{\citenamefont{Kuramoto and Battogtokh}(2002)}]{kuramoto2002}
\bibinfo{author}{\bibfnamefont{Y.}~\bibnamefont{Kuramoto}} \bibnamefont{and}
  \bibinfo{author}{\bibfnamefont{D.}~\bibnamefont{Battogtokh}},
  \bibinfo{journal}{Nonlinear Phenom. Complex Syst.}
  \textbf{\bibinfo{volume}{5}}, \bibinfo{pages}{380} (\bibinfo{year}{2002}).

\bibitem[{\citenamefont{Abrams and Strogatz}(2004)}]{abrams2004}
\bibinfo{author}{\bibfnamefont{D.~M.} \bibnamefont{Abrams}} \bibnamefont{and}
  \bibinfo{author}{\bibfnamefont{S.~H.} \bibnamefont{Strogatz}},
  \bibinfo{journal}{Phys. Rev. Lett.} \textbf{\bibinfo{volume}{93}},
  \bibinfo{pages}{174102} (\bibinfo{year}{2004}).

\bibitem[{\citenamefont{Motter}(2010)}]{motter2010}
\bibinfo{author}{\bibfnamefont{A.~E.} \bibnamefont{Motter}},
  \bibinfo{journal}{Nat. Phys.} \textbf{\bibinfo{volume}{6}},
  \bibinfo{pages}{164} (\bibinfo{year}{2010}).

\bibitem[{\citenamefont{Hagerstrom et~al.}(2012)\citenamefont{Hagerstrom,
  Murphy, Roy, H{\"o}vel, Omelchenko, and Sch{\"o}ll}}]{hagerstrom2012}
\bibinfo{author}{\bibfnamefont{A.~M.} \bibnamefont{Hagerstrom}},
  \bibinfo{author}{\bibfnamefont{T.~E.} \bibnamefont{Murphy}},
  \bibinfo{author}{\bibfnamefont{R.}~\bibnamefont{Roy}},
  \bibinfo{author}{\bibfnamefont{P.}~\bibnamefont{H{\"o}vel}},
  \bibinfo{author}{\bibfnamefont{I.}~\bibnamefont{Omelchenko}},
  \bibnamefont{and}
  \bibinfo{author}{\bibfnamefont{E.}~\bibnamefont{Sch{\"o}ll}},
  \bibinfo{journal}{Nat. Phys.} \textbf{\bibinfo{volume}{8}},
  \bibinfo{pages}{658} (\bibinfo{year}{2012}).

\bibitem[{\citenamefont{Totsuji et~al.}(2001)\citenamefont{Totsuji, Totsuji,
  and Tsuruta}}]{totsuji2001}
\bibinfo{author}{\bibfnamefont{H.}~\bibnamefont{Totsuji}},
  \bibinfo{author}{\bibfnamefont{C.}~\bibnamefont{Totsuji}}, \bibnamefont{and}
  \bibinfo{author}{\bibfnamefont{K.}~\bibnamefont{Tsuruta}},
  \bibinfo{journal}{Phys. Rev. E} \textbf{\bibinfo{volume}{64}},
  \bibinfo{pages}{066402} (\bibinfo{year}{2001}).

\bibitem[{\citenamefont{Zhdanov et~al.}(2011)\citenamefont{Zhdanov, Thoma, and
  Morfill}}]{zhdanov2011spontaneous}
\bibinfo{author}{\bibfnamefont{S.~K.} \bibnamefont{Zhdanov}},
  \bibinfo{author}{\bibfnamefont{M.~H.} \bibnamefont{Thoma}}, \bibnamefont{and}
  \bibinfo{author}{\bibfnamefont{G.~E.} \bibnamefont{Morfill}},
  \bibinfo{journal}{New J. Phys.} \textbf{\bibinfo{volume}{13}},
  \bibinfo{pages}{013039} (\bibinfo{year}{2011}).

\bibitem[{\citenamefont{Ivlev et~al.}(2003)\citenamefont{Ivlev, Konopka,
  Morfill, and Joyce}}]{ivlev2003}
\bibinfo{author}{\bibfnamefont{A.~V.} \bibnamefont{Ivlev}},
  \bibinfo{author}{\bibfnamefont{U.}~\bibnamefont{Konopka}},
  \bibinfo{author}{\bibfnamefont{G.~E.} \bibnamefont{Morfill}},
  \bibnamefont{and} \bibinfo{author}{\bibfnamefont{G.}~\bibnamefont{Joyce}},
  \bibinfo{journal}{Phys. Rev. E} \textbf{\bibinfo{volume}{68}},
  \bibinfo{pages}{026405} (\bibinfo{year}{2003}).

\bibitem[{\citenamefont{R{\"o}cker et~al.}(2014)\citenamefont{R{\"o}cker,
  Ivlev, Zhdanov, and Morfill}}]{rocker2014effect}
\bibinfo{author}{\bibfnamefont{T.~B.} \bibnamefont{R{\"o}cker}},
  \bibinfo{author}{\bibfnamefont{A.~V.} \bibnamefont{Ivlev}},
  \bibinfo{author}{\bibfnamefont{S.~K.} \bibnamefont{Zhdanov}},
  \bibnamefont{and} \bibinfo{author}{\bibfnamefont{G.~E.}
  \bibnamefont{Morfill}}, \bibinfo{journal}{Phys. Rev. E}
  \textbf{\bibinfo{volume}{89}}, \bibinfo{pages}{013104}
  (\bibinfo{year}{2014}).

\bibitem[{\citenamefont{Zhdanov et~al.}(2003)\citenamefont{Zhdanov, Quinn,
  Samsonov, and Morfill}}]{zhdanov2003large}
\bibinfo{author}{\bibfnamefont{S.}~\bibnamefont{Zhdanov}},
  \bibinfo{author}{\bibfnamefont{R.~A.} \bibnamefont{Quinn}},
  \bibinfo{author}{\bibfnamefont{D.}~\bibnamefont{Samsonov}}, \bibnamefont{and}
  \bibinfo{author}{\bibfnamefont{G.~E.} \bibnamefont{Morfill}},
  \bibinfo{journal}{New J. Phys.} \textbf{\bibinfo{volume}{5}},
  \bibinfo{pages}{74} (\bibinfo{year}{2003}).

\bibitem[{\citenamefont{Sheridan}(2008)}]{sheridan2008}
\bibinfo{author}{\bibfnamefont{T.~E.} \bibnamefont{Sheridan}},
  \bibinfo{journal}{Phys. Plasmas} \textbf{\bibinfo{volume}{15}},
  \bibinfo{pages}{103702} (\bibinfo{year}{2008}).

\bibitem[{\citenamefont{Sheridan}(2009)}]{sheridan2009}
\bibinfo{author}{\bibfnamefont{T.~E.} \bibnamefont{Sheridan}},
  \bibinfo{journal}{Phys. Plasmas} \textbf{\bibinfo{volume}{16}},
  \bibinfo{pages}{3705} (\bibinfo{year}{2009}).

\bibitem[{\citenamefont{Schofield}(1973)}]{schofield1973}
\bibinfo{author}{\bibfnamefont{P.}~\bibnamefont{Schofield}},
  \bibinfo{journal}{Comput. Phys. Commun.} \textbf{\bibinfo{volume}{5}},
  \bibinfo{pages}{17} (\bibinfo{year}{1973}).

\bibitem[{\citenamefont{Beeman}(1976)}]{beeman1976}
\bibinfo{author}{\bibfnamefont{D.}~\bibnamefont{Beeman}}, \bibinfo{journal}{J.
  Comput. Phys.} \textbf{\bibinfo{volume}{20}}, \bibinfo{pages}{130}
  (\bibinfo{year}{1976}).

\bibitem[{\citenamefont{Donk{\'o} et~al.}(2008)\citenamefont{Donk{\'o}, Kalman,
  and Hartmann}}]{donko2008}
\bibinfo{author}{\bibfnamefont{Z.}~\bibnamefont{Donk{\'o}}},
  \bibinfo{author}{\bibfnamefont{G.~J.} \bibnamefont{Kalman}},
  \bibnamefont{and} \bibinfo{author}{\bibfnamefont{P.}~\bibnamefont{Hartmann}},
  \bibinfo{journal}{J. Phys. Condens. Matter} \textbf{\bibinfo{volume}{20}},
  \bibinfo{pages}{413101} (\bibinfo{year}{2008}).

\bibitem[{\citenamefont{Ivlev et~al.}(2015)\citenamefont{Ivlev, R\"ocker,
  Cou\"edel, Nosenko, and Du}}]{ivlev2015wave}
\bibinfo{author}{\bibfnamefont{A.~V.} \bibnamefont{Ivlev}},
  \bibinfo{author}{\bibfnamefont{T.~B.} \bibnamefont{R\"ocker}},
  \bibinfo{author}{\bibfnamefont{L.}~\bibnamefont{Cou\"edel}},
  \bibinfo{author}{\bibfnamefont{V.}~\bibnamefont{Nosenko}}, \bibnamefont{and}
  \bibinfo{author}{\bibfnamefont{C.-R.} \bibnamefont{Du}},
  \bibinfo{journal}{Phys. Rev. E} \textbf{\bibinfo{volume}{91}},
  \bibinfo{pages}{063108} (\bibinfo{year}{2015}).

\bibitem[{\citenamefont{Durniak and Samsonov}(2011)}]{durniak2011}
\bibinfo{author}{\bibfnamefont{C.}~\bibnamefont{Durniak}} \bibnamefont{and}
  \bibinfo{author}{\bibfnamefont{D.}~\bibnamefont{Samsonov}},
  \bibinfo{journal}{Phys. Rev. Lett.} \textbf{\bibinfo{volume}{106}},
  \bibinfo{pages}{175001} (\bibinfo{year}{2011}).

\bibitem[{\citenamefont{Durniak et~al.}(2013)\citenamefont{Durniak, Samsonov,
  Ralph, Zhdanov, and Morfill}}]{durniak2013}
\bibinfo{author}{\bibfnamefont{C.}~\bibnamefont{Durniak}},
  \bibinfo{author}{\bibfnamefont{D.}~\bibnamefont{Samsonov}},
  \bibinfo{author}{\bibfnamefont{J.~F.} \bibnamefont{Ralph}},
  \bibinfo{author}{\bibfnamefont{S.}~\bibnamefont{Zhdanov}}, \bibnamefont{and}
  \bibinfo{author}{\bibfnamefont{G.}~\bibnamefont{Morfill}},
  \bibinfo{journal}{Phys. Rev. E} \textbf{\bibinfo{volume}{88}},
  \bibinfo{pages}{053101} (\bibinfo{year}{2013}).

\bibitem[{\citenamefont{Peeters and Wu}(1987)}]{peeters1987}
\bibinfo{author}{\bibfnamefont{F.~M.} \bibnamefont{Peeters}} \bibnamefont{and}
  \bibinfo{author}{\bibfnamefont{X.}~\bibnamefont{Wu}}, \bibinfo{journal}{Phys.
  Rev. A} \textbf{\bibinfo{volume}{35}}, \bibinfo{pages}{3109}
  (\bibinfo{year}{1987}).

\bibitem[{\citenamefont{Dubin}(1997)}]{dubin1997}
\bibinfo{author}{\bibfnamefont{D.~H.~E.} \bibnamefont{Dubin}},
  \bibinfo{journal}{Phys. Rev. E} \textbf{\bibinfo{volume}{55}},
  \bibinfo{pages}{4017} (\bibinfo{year}{1997}).

\bibitem[{\citenamefont{Lifshitz and Pitaevskii}(1981)}]{lifshitz1981}
\bibinfo{author}{\bibfnamefont{E.~M.} \bibnamefont{Lifshitz}} \bibnamefont{and}
  \bibinfo{author}{\bibfnamefont{L.~P.} \bibnamefont{Pitaevskii}},
  \emph{\bibinfo{title}{Physical Kinetics}} (\bibinfo{publisher}{Pergamon,
  Oxford}, \bibinfo{year}{1981}).

\bibitem[{\citenamefont{Frisch}(1995)}]{frisch1995}
\bibinfo{author}{\bibfnamefont{U.}~\bibnamefont{Frisch}},
  \emph{\bibinfo{title}{Turbulence: The Legacy of A. N. Kolmogorov}}
  (\bibinfo{publisher}{Cambridge University Press}, \bibinfo{year}{1995}).

\bibitem[{\citenamefont{Schwabe et~al.}(2014)\citenamefont{Schwabe, Zhdanov,
  R{\"a}th, Graves, Thomas, and Morfill}}]{schwabe2014}
\bibinfo{author}{\bibfnamefont{M.}~\bibnamefont{Schwabe}},
  \bibinfo{author}{\bibfnamefont{S.}~\bibnamefont{Zhdanov}},
  \bibinfo{author}{\bibfnamefont{C.}~\bibnamefont{R{\"a}th}},
  \bibinfo{author}{\bibfnamefont{D.~B.} \bibnamefont{Graves}},
  \bibinfo{author}{\bibfnamefont{H.~M.} \bibnamefont{Thomas}},
  \bibnamefont{and} \bibinfo{author}{\bibfnamefont{G.~E.}
  \bibnamefont{Morfill}}, \bibinfo{journal}{Phys. Rev. Lett.}
  \textbf{\bibinfo{volume}{112}}, \bibinfo{pages}{115002}
  (\bibinfo{year}{2014}).

\bibitem[{\citenamefont{Zhdanov et~al.}(2015)\citenamefont{Zhdanov, Schwabe,
  R{\"a}th, Thomas, and Morfill}}]{zhdanov2015wave}
\bibinfo{author}{\bibfnamefont{S.}~\bibnamefont{Zhdanov}},
  \bibinfo{author}{\bibfnamefont{M.}~\bibnamefont{Schwabe}},
  \bibinfo{author}{\bibfnamefont{C.}~\bibnamefont{R{\"a}th}},
  \bibinfo{author}{\bibfnamefont{H.~M.} \bibnamefont{Thomas}},
  \bibnamefont{and} \bibinfo{author}{\bibfnamefont{G.~E.}
  \bibnamefont{Morfill}}, \bibinfo{journal}{EPL}
  \textbf{\bibinfo{volume}{110}}, \bibinfo{pages}{35001}
  (\bibinfo{year}{2015}).

\bibitem[{\citenamefont{Matveev et~al.}(2010)\citenamefont{Matveev, Andreev,
  and Pustilnik}}]{matveev2010}
\bibinfo{author}{\bibfnamefont{K.~A.} \bibnamefont{Matveev}},
  \bibinfo{author}{\bibfnamefont{A.~V.} \bibnamefont{Andreev}},
  \bibnamefont{and}
  \bibinfo{author}{\bibfnamefont{M.}~\bibnamefont{Pustilnik}},
  \bibinfo{journal}{Phys. Rev. Lett.} \textbf{\bibinfo{volume}{105}},
  \bibinfo{pages}{046401} (\bibinfo{year}{2010}).

\bibitem[{\citenamefont{Yang and Chen}(2003)}]{yang2003}
\bibinfo{author}{\bibfnamefont{B.}~\bibnamefont{Yang}} \bibnamefont{and}
  \bibinfo{author}{\bibfnamefont{G.}~\bibnamefont{Chen}},
  \bibinfo{journal}{Phys. Rev. B} \textbf{\bibinfo{volume}{67}},
  \bibinfo{pages}{195311} (\bibinfo{year}{2003}).

\bibitem[{\citenamefont{Maurel et~al.}(2004)\citenamefont{Maurel, Mercier, and
  Lund}}]{maurel2004}
\bibinfo{author}{\bibfnamefont{A.}~\bibnamefont{Maurel}},
  \bibinfo{author}{\bibfnamefont{J.-F.} \bibnamefont{Mercier}},
  \bibnamefont{and} \bibinfo{author}{\bibfnamefont{F.}~\bibnamefont{Lund}},
  \bibinfo{journal}{J. Acoust. Soc. Am.} \textbf{\bibinfo{volume}{115}},
  \bibinfo{pages}{2773} (\bibinfo{year}{2004}).

\bibitem[{\citenamefont{Maurel et~al.}(2006)\citenamefont{Maurel, Pagneux,
  Boyer, and Lund}}]{maurel2006}
\bibinfo{author}{\bibfnamefont{A.}~\bibnamefont{Maurel}},
  \bibinfo{author}{\bibfnamefont{V.}~\bibnamefont{Pagneux}},
  \bibinfo{author}{\bibfnamefont{D.}~\bibnamefont{Boyer}}, \bibnamefont{and}
  \bibinfo{author}{\bibfnamefont{F.}~\bibnamefont{Lund}},
  \bibinfo{journal}{Proc. R. Soc. A} \textbf{\bibinfo{volume}{462}},
  \bibinfo{pages}{2607} (\bibinfo{year}{2006}).

\bibitem[{\citenamefont{Lugueva and Luguev}(2004)}]{lugueva2004}
\bibinfo{author}{\bibfnamefont{N.~V.} \bibnamefont{Lugueva}} \bibnamefont{and}
  \bibinfo{author}{\bibfnamefont{S.~M.} \bibnamefont{Luguev}},
  \bibinfo{journal}{High Temp} \textbf{\bibinfo{volume}{42}},
  \bibinfo{pages}{54} (\bibinfo{year}{2004}).

\bibitem[{\citenamefont{Knapek et~al.}(2007)\citenamefont{Knapek, Samsonov,
  Zhdanov, Konopka, and Morfill}}]{knapek2007}
\bibinfo{author}{\bibfnamefont{C.~A.} \bibnamefont{Knapek}},
  \bibinfo{author}{\bibfnamefont{D.}~\bibnamefont{Samsonov}},
  \bibinfo{author}{\bibfnamefont{S.}~\bibnamefont{Zhdanov}},
  \bibinfo{author}{\bibfnamefont{U.}~\bibnamefont{Konopka}}, \bibnamefont{and}
  \bibinfo{author}{\bibfnamefont{G.~E.} \bibnamefont{Morfill}},
  \bibinfo{journal}{Phys. Rev. Lett.} \textbf{\bibinfo{volume}{98}},
  \bibinfo{pages}{015004} (\bibinfo{year}{2007}).

\bibitem[{\citenamefont{Knapek et~al.}(2013)\citenamefont{Knapek, Durniak,
  Samsonov, and Morfill}}]{knapek2013}
\bibinfo{author}{\bibfnamefont{C.~A.} \bibnamefont{Knapek}},
  \bibinfo{author}{\bibfnamefont{C.}~\bibnamefont{Durniak}},
  \bibinfo{author}{\bibfnamefont{D.}~\bibnamefont{Samsonov}}, \bibnamefont{and}
  \bibinfo{author}{\bibfnamefont{G.~E.} \bibnamefont{Morfill}},
  \bibinfo{journal}{Phys. Rev. Lett.} \textbf{\bibinfo{volume}{110}},
  \bibinfo{pages}{035001} (\bibinfo{year}{2013}).

\bibitem[{\citenamefont{Ling}(2005)}]{ling2005}
\bibinfo{author}{\bibfnamefont{X.~S.} \bibnamefont{Ling}},
  \bibinfo{journal}{Nat. Mater.} \textbf{\bibinfo{volume}{4}},
  \bibinfo{pages}{360} (\bibinfo{year}{2005}).

\bibitem[{\citenamefont{Kittel}(1976)}]{kittel1976}
\bibinfo{author}{\bibfnamefont{C.}~\bibnamefont{Kittel}},
  \emph{\bibinfo{title}{Introduction to Solid State Physics}}
  (\bibinfo{publisher}{Wiley, New York}, \bibinfo{year}{1976}),
  \bibinfo{edition}{5th} ed.

\bibitem[{\citenamefont{Paulose et~al.}(2015)\citenamefont{Paulose, Chen, and
  Vitelli}}]{paulose2015}
\bibinfo{author}{\bibfnamefont{J.}~\bibnamefont{Paulose}},
  \bibinfo{author}{\bibfnamefont{B.~G.} \bibnamefont{Chen}}, \bibnamefont{and}
  \bibinfo{author}{\bibfnamefont{V.}~\bibnamefont{Vitelli}},
  \bibinfo{journal}{Nat. Phys.} \textbf{\bibinfo{volume}{11}},
  \bibinfo{pages}{153} (\bibinfo{year}{2015}).

\bibitem[{\citenamefont{Peierls}(1929)}]{peierls1929}
\bibinfo{author}{\bibfnamefont{R.}~\bibnamefont{Peierls}},
  \bibinfo{journal}{Annalen der Physik} \textbf{\bibinfo{volume}{395}},
  \bibinfo{pages}{1055} (\bibinfo{year}{1929}).

\bibitem[{\citenamefont{Berman et~al.}(1951)\citenamefont{Berman, Simon, and
  Wilks}}]{berman1951}
\bibinfo{author}{\bibfnamefont{R.}~\bibnamefont{Berman}},
  \bibinfo{author}{\bibfnamefont{F.~E.} \bibnamefont{Simon}}, \bibnamefont{and}
  \bibinfo{author}{\bibfnamefont{J.}~\bibnamefont{Wilks}},
  \bibinfo{journal}{Nature (London)} \textbf{\bibinfo{volume}{168}},
  \bibinfo{pages}{277} (\bibinfo{year}{1951}).

\bibitem[{\citenamefont{Han}(1996)}]{han1996}
\bibinfo{author}{\bibfnamefont{Y.-J.} \bibnamefont{Han}},
  \bibinfo{journal}{Phys. Rev. B} \textbf{\bibinfo{volume}{54}},
  \bibinfo{pages}{8977} (\bibinfo{year}{1996}).

\bibitem[{\citenamefont{Bloembergen}(1965)}]{bloembergen1965}
\bibinfo{author}{\bibfnamefont{N.}~\bibnamefont{Bloembergen}},
  \emph{\bibinfo{title}{Nonlinear Optics: A Lecture Note and Reprint Volume}}
  (\bibinfo{publisher}{Benjamin inc. (New York)}, \bibinfo{year}{1965}).

\end{thebibliography}

\end{document}